\documentclass[a4paper,single column,12pt]{article}
\usepackage{epsfig,amsmath,amssymb}
\usepackage[colorlinks=true,linktocpage=true,linkcolor=blue,citecolor=blue]{hyperref}
\usepackage{graphicx}
\usepackage{dcolumn}
\usepackage{bm}
\usepackage{subfig}
\usepackage{float}
\usepackage{tabularx}
\usepackage{hhline}
\usepackage{color}
\newcommand{\old}[1]{}

\newcommand{\be}{\begin{equation}}
\newcommand{\ee}{\end{equation}}
\newcommand{\ba}{\begin{eqnarray}}
\newcommand{\ea}{\end{eqnarray}}
\newcommand{\bi}{\begin{itemize}}
\newcommand{\ei}{\end{itemize}}

\usepackage{caption}
\begin{document}
\begin{flushright}
{\normalsize
}
\end{flushright}
\vskip 0.1in
\begin{center}
{\large {\bf Landau Damping in a strong magnetic field: Dissociation of Quarkonia }}
\end{center}
\vskip 0.1in
\begin{center}
Mujeeb Hasan$^\dag$\footnote{hasan.dph2014@iitr.ac.in}, Binoy Krishna Patra$^\dag$
\footnote{binoyfph@iitr.ac.in}, Bhaswar
Chatterjee$^\dag$\footnote{bhaswar.mph2016@iitr.ac.in}, and Partha Bagchi$^\ast$\footnote{p.bagchi@vecc.gov.in}
\vskip 0.02in
{\small {\it $^\dag$ Department of Physics, Indian Institute of
Technology Roorkee, Roorkee 247 667, India\\
$^\ast$ Variable Energy Cyclotron Centre, 1/AF Bidhannagar, Kolkata 700 064, 
India} }
\end{center}
\vskip 0.01in
\addtolength{\baselineskip}{0.4\baselineskip} 
\section* {Abstract}
In this article we have investigated the effects of strong magnetic field 
on the properties of quarkonia immersed in a thermal medium of quarks 
and gluons and studied its quasi-free dissociation due to the 
Landau-damping. Thermalizing the Schwinger propagator 
in the lowest Landau levels for quarks and the Feynman propagator for
gluons in real-time formalism, we have calculated the resummed retarded 
and symmetric propagators, which in turn give the real and imaginary
components of dielectric permittivity, respectively. Thus 
the effect of a strongly magnetized hot QCD medium have been encrypted
into the real and imaginary parts of heavy quark interaction
in medium, respectively. The magnetic field affects the large-distance 
interaction more than the short-distance interaction, as a result, the 
real part of potential becomes more attractive and 
the magnitude of imaginary part too becomes larger, compared
to the thermal medium in absence of strong magnetic field.
As a consequence the average size of $J/\psi$'s  and $\psi^\prime$'s are
increased but $\chi_c$'s get shrunk. Similarly the magnetic field 
affects the binding of $J/\psi$'s and $\chi_c$'s discriminately, {\em i.e.}
it decreases the binding of $J/\psi$ and increases for 
$\chi_c$. However, the further increase in magnetic field results in the 
decrease of binding energies. On contrary the 
magnetic field increases the width of the resonances, unless
the temperature is sufficiently high. We have finally studied how the 
presence of
magnetic field affects the dissolution of quarkonia in a thermal
medium due to the Landau damping, where the dissociation temperatures are 
found to increase compared to the thermal medium in absence of magnetic field.
However, further increase of magnetic field decreases 
the dissociation temperatures. {\em For example}, $J/\psi$'s and $\chi_c$'s 
are dissociated at higher temperatures
at 2 $T_c$ and 1.1 $T_c$ at a magnetic field $eB \approx 6~{\rm{and}}~4~ 
m_\pi^2$, respectively, compared to the 
values 1.60 $T_c$ and 0.8 $T_c$ in the absence of magnetic field, respectively.

\noindent PACS:~~ 12.39.-x,11.10.St,12.38.Mh,12.39.Pn
12.75.N, 12.38.G \\
\vspace{1mm}
\noindent{\bf Keywords}: Thermal QCD, Retarded Propagator, Symmetric
Propagator, Dielectric Permittivity, heavy quark potential\\

\section{Introduction}
Quantum Chromodynamics  (QCD) predicts that at sufficiently high temperatures
and/or densities the quarks and gluon confined inside the hadrons 
are liberated into a medium of quarks and gluons, known
as Quark-gluon Plasma (QGP). Over the decades a large number
of activities have been directed towards the production and identification
of this new state of matter theoretically and experimentally
in ultra relativistic heavy-ion collisions (URHIC) with the increasing center
of mass energies ($\sqrt{s}$) at BNL AGS, CERN SPS, BNL RHIC, and CERN LHC
experiments. However, for the non-central events in the above URHICs, a
very strong magnetic field is generated at the very early stages of the
collisions due to very high relative velocities of the spectator quarks
with respect to the fireball~\cite{Skokov:IJMPA'2009,Voronyuk:PRC83'2011}.
Depending on the centralities of the collisions, the strength of the magnetic 
fields may vary from $m_{\pi}^2$ ($\sim 10^{18}$ Gauss) at RHIC to
10 $m_{\pi}^2$ at LHC. However, at extreme cases the magnetic
field may even reach 50 $m_{\pi}^2$ at LHC and even much larger 
values $\sim 10^{5}~m_\pi^2$ in the early universe 
during electroweak phase transition~\cite{Vachaspati:PLB265'1991}. Naive
classical estimates of the lifetime of these
magnetic fields show that it only exists for a
small fraction of the lifetime of QGP. However depending on the transport 
properties
of the plasma the magnetic field may remain strong during the lifetime of 
QGP~\cite{Tuchin:PRC83'2010}.

One particularly suited probe to infer the properties of 
nuclear matter under extreme conditions is 
the heavy-quarkonia. The heavy quark and antiquark ($Q \bar Q$) pairs are 
produced in URHICs on a very short time-scale $t_{\rm prod}\sim 1/2m_{Q}$.
Subsequently they develop into a physical resonance over a formation time
$t_{\rm form}\sim 1/E_{\rm bind}$ ($E_{\rm{bind}}$ is the binding energy of the 
state). They traverse the plasma and later the hadronic matter
before decaying into the dilepton, which is 
eventually detected. This long journey is fairly `hazardous' for the 
quarkonium because even before the formation 
of resonances, the cold nuclear matter may
dissociate the nascent $Q \bar Q$ pairs. However, even after 
the resonances are formed, they react to the presence of a thermal 
medium with the smaller binding energies. Since the mass of 
the charm or bottom quarks is larger than the temperature 
of QGP created in current 
heavy-ion collisions, {\em viz.} $T_{\rm LHC}\leq 0.6$ GeV 
heavy quarkonium bound states may survive 
while traversing the collision center. In that 
process they accumulate information about their 
environment which is imprinted on their depleted production 
yields, which may open up a direct window on the vital properties of
the deconfined medium, {\em namely} the temperature and the 
presence of strong magnetic fields. Therefore the goal of the present work
is to understand theoretically the properties of heavy 
quarkonium under realistic conditions existing in an environment at 
high temperatures in the presence of strong magnetic fields.

Our understanding of heavy quarkonium has made a significant step forward 
with the computations of effective field theories (EFT) from the
underlying theory - QCD, {\em such as} 
non-relativistic QCD (NRQCD) and potential NRQCD, 
which are synthesized by separating the intrinsic scales 
of heavy quark bound states ({\em e.g.} mass, velocity, binding energy) 
as well as the additional scales of thermal medium ({\em e.g.} $T$, $g$T, 
$g^2$T) in weak-coupling regime, in overall comparison with $\Lambda_{QCD}$.
However, the separation of scales in EFT is not always evident in 
realistic conditions achieved at URHICs, so one needs the first-principle 
lattice QCD simulations to study the quarkonia in a medium
even without the potential models rather by 
the spectral functions in terms of the Euclidean meson correlation 
functions~\cite{Alberico:PRD77'2008}. However the reconstruction of the 
spectral functions turns out to be very difficult because the temporal extent 
decreases at large temperature. Thereby the studies of quarkonia using
the potential models at finite temperature complement the lattice studies.

For a long time phenomenological potential models had been deployed 
in the literature, which were not based on the systematic derivations 
from QCD. The color singlet free energies extracted 
from the correlation function of Polyakov loops, 
which is computed from the first-principle lattice
QCD simulations, has been commonly advocated as an appropriate potential
to study the quarkonia in vacuum as well as in 
medium. The perturbative computations of the 
potential at high temperatures show that the $Q \bar Q$ potential becomes 
complex, where the real part gets screened due to the presence
of deconfined color charges~\cite{Matsui:PLB178'1986} and the imaginary-part
~\cite{Escobedo:PRA78'2008,Brambilla:PRD78'2008,Laine:JHEP0703'2007,
Beraudo:NPA806'2008} attributes the thermal width of the resonance.
The physics of quarkonium dissociation in a medium
has been refined over the last decade, where the resonances
were initially thought to be dissociated when the screening becomes 
sufficiently strong, the potential becomes too weak to hold $Q \bar Q$ 
together. Nowadays the dissociation is thought to be 
mainly due to the broadening of the width of resonances in a medium.
The broadening arises mainly either by the inelastic parton
scattering process mediated by the spacelike gluons, known as Landau damping
~\cite{Laine:JHEP0703'2007} or due to the gluo-dissociation process in which
the color singlet state undergoes into a color octet state by a 
hard thermal gluon~\cite{Brambilla:JHEP1305'2013}. The later processes 
becomes dominant when the temperature of medium is smaller than the binding 
energy of the particular resonance. Recently one of us estimated the imaginary 
component of the potential perturbatively in resummed thermal field theory,
where the inclusion of a confining string term makes the
(magnitude) imaginary component smaller~\cite{Lata:PRD89'2014}
, compared to
the medium modification of the perturbative term alone
\cite{Adiran:PRD79'2009}. Even in strong coupling limit the potential 
extracted through AdS/CFT correspondence develops
an imaginary component beyond a critical separation of $Q \bar Q$ 
pair~\cite{Binoy:PRD92'2015,Binoy:PRD91'2015}.
In a similar calculation, generalized Gauss law relates the 
numerically simulated values of the potential to the in-medium permittivity 
of the
QCD medium conventionally parameterized by the so called Debye mass
pair~\cite{Rothkopf:PRL'2012}.

The discussions referred above were limited for the 
simplest possible setting in heavy-ion
phenomenology for fully central collisions but most events occur 
with a finite impact parameter where an extremely large magnetic fields
may be produced. Recently some of us have explored the effects of
strong magnetic field on the properties of heavy-quarkonium 
by computing the real part of $Q \bar Q$ potential \cite{Mujeeb:EPJC77'2017}
as well as on the QCD thermodynamics~\cite{SRath:JHEP1712'2017}. 
However, such purely real potential alone 
cannot capture the physics relevant for in-medium modification 
of quarkonium states so we aim to estimate the imaginary component of
the potential perturbatively in the real-time formalism and investigate
how the properties of quarkonia in a thermal QCD medium get affected
by the presence of strong magnetic field. Recently there was an attempt
to derive the complex heavy quark potential due to an external strong
magnetic field in a generalised Gauss law \cite{Balbeer:1711'2017}, where
the imaginary part of in-medium permittivity, 
$\epsilon (k)$ is heuristically obtained by simply replacing the Debye mass 
in the absence of magnetic field by the same in the presence of
strong magnetic field. In our calculation, we aim to calculate meticulously 
the imaginary part of retarded gluon self-energy due to quark loop and 
gluon loop separately, similar to the calculation of the real part. It is 
found that the imaginary part due to quark loop is proportional to the square 
of the quark masses and does not depend on the temperature directly (apart
from the Debye mass). As a result, the momentum dependence will be completely
different from their calculation \cite{Balbeer:1711'2017}, which can be 
understood
by the dimensional reduction caused by the effect of magnetic field 
to quark dynamics, {\em not} the gluon dynamics.

Our work proceeds as follows. First we calculate the resummed retarded/advanced 
and symmetric gluon propagator by calculating the real and imaginary part 
of retarded/advanced gluon self-energies for a thermal QCD medium in 
the presence of strong magnetic field in subsections 2.1 and 2.2, respectively.
Next the real and imaginary component of dielectric permittivities are
obtained by taking the static limit of the resummed retarded and symmetric 
propagators, whose inverse Fourier transform gives 
the real and imaginary parts of heavy quark potential in the coordinate space 
in subsection 3.1 and 3.2, respectively. The real part of potential is thereafter solved numerically by 
the Schr\"{o}dinger equation to obtain both the energy eigenvalues and 
eigenfunctions to calculate the size and binding energies of quarkonia 
in subsection 4.1. In Section 4.2 we deals with the imaginary component 
in a time-independent perturbation theory to estimate the medium-induced 
thermal width of the resonances, which facilitates to study the dissociation
due to the Landau damping. Finally we will conclude in Section 5.

\section{The resummed gluon propagator in strong magnetic field}
In Keldysh representation of real-time formalism, the retarded (R), 
advanced (A) and symmetric (S) propagators are written as the linear
combination of the components of matrix propagator:
\begin{eqnarray}
\label{2a6}
   D_R^0 = D_{11}^0 - D_{12}^0 ~,~ D_A^0 = D_{11}^0 - D_{21}^0 ~,~
   D_S^0 = D_{11}^0 + D_{22}^0~.
\end{eqnarray}
Similar representation for self-energies can also be worked out in terms 
of components of self-energy matrix through the retarded ($\Pi_R$), 
advanced ($\Pi_A$) and symmetric ($\Pi_S$) self energies. 

The resummation for the above propagators is done by the Dyson-Schwinger 
equation. For the static potential, we need only the 
temporal (longitudinal) component of the propagator and its evaluation is easier in the 
Coulomb gauge so the temporal component of retarded/advanced propagator
is resummed as
\begin{eqnarray}
 D^L_{R,A}=D^{L(0)}_{R,A}+D^{L(0)}_{R,A}\Pi^L_{R,A}{D}^L_{R,A}~, \label{3b2}
\end{eqnarray}
whereas the resummation for symmetric propagator is done as 
\begin{equation}
D^L_{S}=D^{L(0)}_{S}+D^{L(0)}_{R} \Pi^L_R D^L_{S(0)}+D_S^0\Pi_{A} {D}_{A}+
D_{R}^0\Pi _{S}{D}_{A}~.
\label{symmetric}
\end{equation}
Thus the resummed retarded (advanced) and symmetric propagators
can be expressed explicitly by the self-energies as
\begin{eqnarray}
D^{L}_{R,A}(k)&=&\frac{1}{{\textbf{k}}^2-\rm{Re} \Pi^{L}_{R}(k)\mp i \rm{Im} 
\Pi^{L}_{R}(k)},\label{lon_ret} \\
D^{L}_{S}(k)&=&(1+2n_{B}(k_0))~{\rm{sgn}}(k_0) \left(D^{L}_{R}(k)-D^{L}_{A}(k)
\right),
\label{lon_sym}
\end{eqnarray}
where the factor, $(1+2n_{B}(k_0)) {\rm{sgn}} (k_0)$ and the difference, 
$\left(D^{L}_{R}(k)-D^{L}_{A}(k)\right)$ can be obtained as
~\cite{Adiran:PRD79'2009,Magaret:EPJC7'1999}
\begin{eqnarray}
(1+2n_{B}(k_0)) {\rm{sgn}} (k_0) &=& \frac{2T}{k_0}, \label{factor}\\
\left(D^{L}_{R}(k)-D^{L}_{A}(k)\right)&=&\frac{2i\rm{Im} \Pi^{L}_{R}(k)}
{\big[\textbf{k}^2-\rm{Re} \Pi^{L}_{R}(k)\big]^2+\big[\rm{Im} \Pi^{L}_{R}
(k)\big]^2},
\label{retarded_advanced}
\end{eqnarray}
with the following identities
\begin{eqnarray}
\rm{Re}\Pi^{L}_{R}(k) &=&\rm{Re}\Pi^{L}_{A}
(k),\\
\rm{Im} \Pi^{L}_{R}(k)&=&-\rm{Im} \Pi^{L}_{A}(k).
\end{eqnarray}
It is thus learnt that only the real and imaginary parts of the 
longitudinal component of retarded 
self-energy suffice to calculate the resummed retarded, advanced
and symmetric propagator in a strongly magnetized hot QCD medium.

For calculating the retarded self-energy, we 
need to evaluate the matrix propagator in a thermal medium in the presence 
of strong magnetic field for quarks and gluons. The magnetic field affects 
only the quark propagator {\em via} the projection operator and its 
dispersion relation. So we will now revisit the vacuum quark 
propagator in a strong magnetic field and then thermalize it
in a real-time formalism, which in turn computes the gluon
self-energy for the quark-loop diagram.
We start with the vacuum quark propagator in coordinate-space, using 
the Schwinger's proper-time method \cite{Schwinger:PR82'1951} 
\begin{equation}\label{S(X,Y)}
S(y,y^\prime)=\phi(y,y^\prime)\int\frac{d^4p}{(2\pi)^4}e^{-ip(y-y^\prime)}
S(p)~,
\end{equation}
where the phase factor, $\phi(y,y^\prime)$ defined by
\begin{equation}
\phi(y,y^\prime)=e^{i|q_f|\int^{y}_{y^\prime} A^\mu(\zeta)d\zeta_\mu}.
\end{equation}
is a gauge-dependent quantity, which is responsible for breaking of 
translational invariance. For a single fermion line, it is possible to 
gauge away the phase factor by an appropriate gauge transformation 
for a symmetric gauge in a magnetic field directed along the $z$ axis.
Thus one can express the propagator in the momentum-space
\cite{Tsai:PRD10'1974,Chyi:PRD62'2000} as an integral over the proper-time 
($s$)
\begin{equation}
iS(p)=\int_0^\infty \frac{1}{eB}\frac{ds}{\cos(s)}
e^{-is\left[m^2_f-p_{\|}^2+\frac{\tan(s)}{s}
p_\bot^2\right]}
\left[(\cos(s) +\gamma_{1}\gamma_{2} 
\sin(s))(m_f+\gamma\cdot p_{\|})
-\frac{\gamma\cdot p_{\bot}}{\cos (s)}\right],
\label{sch_mom}
\end{equation}
which can be expressed more conveniently in a discrete form by the 
associated Laguerre polynomials
\begin{equation}
iS_n (p)=\sum_n\frac{-id_n(\alpha)D+d^\prime_n(\alpha)
\bar{D}}{p_L^2+2neB}
+i\frac{\gamma\cdot p_\bot}{p_\bot^2}~,
\label{prop_lag}
\end{equation}
with the notations in Ref\cite{Chyi:PRD62'2000}.

In a strong magnetic field (SMF) limit both parallel and perpendicular 
components of quark momenta are smaller than the magnetic field 
(i.e. $p^2_\parallel,p^2_\perp\ll|q_fB| \gg T^2$) so the transitions to the 
higher Landau levels ($n\geq1$) are suppressed.  Therefore only the lowest 
Landau levels (LLL) are populated, hence
the vacuum propagator for quarks in the momentum-space for LLL ($n=0$) becomes 
\begin{equation}
iS_0(p)=\frac{(1+\gamma^{0}\gamma^{3}\gamma^{5})
(\gamma^{0}p_{0}-\gamma^{3}p_{z}+m_f)}
{p_{\parallel}^2-m^2_f+i\epsilon} e^{-\frac{p_{\perp}^2}
{\mid q_fB\mid}},
\label{vacrop}
\end{equation}
where $m_f$ and $q_f$ are the mass and electric charge of $f^{th}$ flavour, 
respectively. However, in real-time formalism, the propagator in 
a thermal medium acquires a ($2\times{2}$) matrix structure~\cite{Magaret:EPJC7'1999}  
\begin{equation}
S(p) =
\begin{pmatrix}
S_0(p)+n_F(p_0) (S^\ast_0 (p) -S_0(p)) & 
\sqrt{n_F(p_0) (1-n_F(p_0))} (S^\ast_0 (p) -S_0(p)) \\
-\sqrt{n_F(p_0) (1-n_F(p_0)} (S^\ast_0 (p) -S_0(p)) &
-S^\ast_0(p)+n_F(p_0) (S^\ast_0 (p) -S_0(p)) 
\end{pmatrix}~,
\label{mag_prop}
\end{equation}
where $n_F(p_0)$ is the quark distribution function.
Thus, the $11$- and $12$-components can be read off
\begin{eqnarray}
iS_{11}(p)&=&\Bigg[\frac{1}{{p_{\parallel}^2-m_f^2+
i\epsilon}}+2\pi i n_F(p_0)\delta(p_{\parallel}^2-m_f^2)\Bigg]
(1+\gamma^{0}\gamma^{3}\gamma^{5})(\gamma^{0}p_{0}-\gamma^{3}
p_{z}+m_f) e^{\frac{-p_{\perp}^2}{\mid q_fB \mid}},
\label{propagator_11}\\
S_{12}(p)&=&-2\pi\sqrt{n_F(p_0)(1-n_F(p_0))}
\delta(p_{\parallel}^2-m_f^2)(1+\gamma^{0}\gamma^{3}\gamma^{5})
(\gamma^{0}p_{0}-\gamma^{3}p_{z}+m_f)e^{\frac{-p_{\perp}^2}
{\mid q_fB \mid}}.
\label{propagator_12}
\end{eqnarray}

However, for gluons, the form of the vacuum propagator remains unaffected
by the magnetic field, {\em i.e.} 
\begin{eqnarray}\label{1 G.P.}
D^{\mu\nu}_0(p)=\frac{ig^{\mu\nu}}{p^2+i\epsilon}
~.\end{eqnarray}
Similar to thermalization of quark propagator, 
the gluon propagator at finite temperature also takes the matrix structure 
in the real-time formalism~\cite{Magaret:EPJC7'1999}  
in terms of the gluon distribution function, $n_B(p_0)$
\begin{equation}
D^{\mu \nu}(p) =
\begin{pmatrix}
D^{\mu \nu}_0(p)+n_B(p_0) (D^{\ast \mu \nu}_0 (p) +D^{\mu \nu}_0(p)) & 
\sqrt{n_B(p_0) (1+n_B(p_0))} (D^{\ast \mu \nu}_0 (p) +D^{\mu \nu}_0(p)) \\
\sqrt{n_B(p_0) (1+n_B(p_0))} (D^{\ast \mu \nu}_0 (p) +D^{\mu \nu}_0(p)) &
D^{\ast \mu \nu}_0(p)+n_B(p_0) (D^{\ast \mu \nu}_0 (p) +D^{\mu \nu}_0(p)) 
\end{pmatrix}.
\label{temp_prop}
\end{equation}
The above matrices (\ref{mag_prop}, \ref{temp_prop}) will be used to calculate the 
retarded/advanced and symmetric self energies due to quark loop and 
gluon loops, respectively in the next section.

\subsection{Real part of retarded gluon self energy in real-time formalism}
In Keldysh representation of real-time formaism, the evaluation of the real 
part of retarded gluon self-energy requires only the 
real part of 11-component of self-energy matrix
\begin{equation}
{\rm{Re}} \Pi_R(k)={\rm{Re}} \Pi_{11}(k). 
\end{equation}

There are four Feynman diagrams, {\em e.g.} tadpole, gluon-loop, 
ghost-loop and quark-loop, which contribute to the gluon self-energy.
Since only the quark-loop diagram is affected by the presence of the 
magnetic field in the thermal medium so we first calculate the 
quark-loop in SMF limit and then obtain the thermal contributions 
due to the remaining gluon-loop diagrams.

Using the matrix propagator (\ref{mag_prop}) for quarks in real-time
formalism, the $11$-component 
of the gluon self-energy matrix for the quark-loop (omitting the prefix
$11$) can be written as 
{\small{
\begin{eqnarray}
\nonumber\Pi^{\mu\nu}(k) &=& i\frac{g^2}
{2}\sum_f\int\frac{{d^2p_\perp}{d^2p_\parallel}}{(2\pi)^4} {\rm{Tr}}\left[\gamma^\mu
(1+\gamma^{0}\gamma^{3}\gamma^{5})\left(\gamma^0p_0-\gamma^3p_z+m_f\right)\gamma^\nu
(1+\gamma^{0}\gamma^{3}\gamma^{5})\left(\gamma^0q_0-\gamma^3q_z+m_f\right)\right] 
\nonumber\\ && 
\times\left[\frac{1}{p^2_\parallel-m^2_f+i\epsilon}+
2\pi{i}n_F\left(p_0\right)\delta\left(p^2_\parallel-m^2_f\right)\right]
e^{-\frac{p^2_\perp}{|q_fB|}}\nonumber \\ 
&& \times\left[\frac{1}{q^2_\parallel-m^2_f+i\epsilon}+
2\pi{i}n_F\left(q_0\right)\delta\left(q^2_\parallel-m^2_f\right)
\right]e^{-\frac{q^2_\perp}{|q_fB|}},
\end{eqnarray}
}}
where the factor $1/2$ arises due to the trace in color space and 
the momentum, $(p+k)$ is replaced by $q$. Here we use the one-loop running 
QCD coupling ($g=\sqrt{4 \pi \alpha_s (eB) }$), which, in strong magnetic field limit, runs exclusively 
with the magnetic field because the most dominant
scale for quarks is no more the temperature of the medium rather
the scale associated with the strong magnetic field. This is exactly
Ferrar et. al  has recently explored the dependence of running coupling
on the magnetic field only by decomposing the momentum
into parallel and perpendicular to the magnetic field~\cite{Ferrer:PRD91_2015}.

Since the momentum integration is factorizable 
into parallel and perpendicular components with respect to the direction 
of magnetic field therefore the component, which depends only the 
transverse momentum, is given by 
\begin{eqnarray}\label{P.C.G.S.E.}
\Pi_\perp (k_\perp) =\frac{\pi|q_fB|}{2}
e^{-\frac{k^2_\perp}{2|q_fB|}}~,
\end{eqnarray}
and the self energy, which depends only the parallel component of 
the momentum, $\Pi^{\mu\nu} 
(k_\parallel)$ is decomposed into vacuum and medium contributions
\begin{eqnarray}
\label{G.S.E.S.M.F.A.}
\Pi_{\parallel}^{\mu\nu}(k_\parallel) 
\equiv \Pi^{\mu\nu}_{\rm vacuum} (k_\parallel)+\Pi^{\mu\nu}_n(k_\parallel)
+\Pi^{\mu\nu}_{n^2}(k_\parallel).
\end{eqnarray}
The vacuum and medium contributions 
having the linear and quadratic dependence on the distribution function,
respectively are given by 
\begin{eqnarray}
\Pi^{\mu \nu}_{\rm{vacuum}}(k_{\parallel}) &=& \frac{ig^2}{2(2\pi)^4}
\int dp_0 dp_z L^{\mu\nu}\left[\frac{1}{(p_{\parallel}^2
-m_f^2+i\epsilon) (q_{\parallel}^2-m_f^2+i\epsilon)}\right],
\label{G.S.E.V.} \\
\Pi^{\mu \nu}_n(k_{\parallel}) &=& -\frac{g^2}
{2(2\pi)^3}\int dp_0 dp_z L^{\mu\nu}\left[n_F(p_0) \frac{
\delta(p_{\parallel}^2-m_f^2)}{(q_{\parallel}^2
-m_f^2+i\epsilon)}
+n_F(q_0) \frac{\delta(q_{\parallel}^2-m_f^2)}
{(p_{\parallel}^2-m_f^2+i\epsilon)}\right],
\label{G.S.E.S.D.} \\
\Pi^{\mu \nu}_{n^2} (k_{\parallel}) &=& -\frac{ig^2}{2(2\pi)^2}
\int dp_0 dp_z L^{\mu\nu}\left[n_{F}(p_0)n_{F}(q_0)
\delta(p_{\parallel}^2-m_f^2)
\delta(q_{\parallel}^2-m_f^2)\right]~,
\label{G.S.E.D.D.}
\end{eqnarray}
where the trace over $\gamma$-matrices, $L^{\mu \nu}$ is 
\begin{eqnarray}
L^{\mu\nu}=8\left[p^\mu_\parallel\cdot{q^\nu_\parallel}
+p^\nu_\parallel\cdot{q^\mu_\parallel}-g^{\mu\nu}_\parallel\left
(p^\mu_\parallel\cdot{q}_{\parallel\mu}
-m^2_f\right)\right]
~.\end{eqnarray}
Now we calculate the real part of the vacuum contribution (\ref{G.S.E.V.}) 
as~\cite{Mujeeb:EPJC77'2017}
\begin{equation}
{\rm{Re}}~\Pi^{\mu\nu}_{\rm{vacuum}}(k_\parallel)=\left(g_{\parallel}^{\mu\nu}
-\frac{k_{\parallel}^{\mu}k_{\parallel}^{\nu}}{k_{\parallel}^2}
\right)\Pi(k_\parallel^2),
\label{self_vacuum}
\end{equation}
where the form factor, $\Pi (k_\parallel^2)$ is given by
\begin{eqnarray}
\Pi (k_\parallel^2)=\frac{g^2}{2\pi^3}\sum_{f} \left[\frac{2m_{f}^2}
{k_{\parallel}^2}
\left(1-\frac{4m_{f}^2}{k_{\parallel}^2}\right)^{-1/2}
\ln \left\lbrace \frac{{\Big(1-\frac{4m_{f}^2}
{k_{\parallel}^2}\Big)}^{1/2}+1}
{{\Big(1-\frac{4m_{f}^2}{k_{\parallel}^2}
\Big)}^{1/2}-1} \right\rbrace +1\right].
\end{eqnarray}
Thus multiplying the transverse momentum dependent part (\ref{P.C.G.S.E.}) to 
the parallel momentum dependent component (\ref{self_vacuum}) and taking the 
static limit ($k_0=0$, $k_x, k_y, k_z \rightarrow 0$),
the longitudinal component of the vacuum part in the limit of
massless flavours becomes  
\begin{equation}
{\rm{Re}}~\Pi^L_{\rm{vacuum}} =-\frac{g^2}{4\pi^2}\sum_{f}|q_{f}B|~,
\label{Massless case}
\end{equation}
whereas for the physical quark masses, it vanishes 
\begin{equation}
\rm{Re}~\Pi^L_{\rm{vacuum}} = 0.
~\label{Massive_case}
\end{equation}

Next the real part of the thermal contribution 
having linear dependence on the distribution function in static limit 
for the massless quarks can be obtained~\cite{Mujeeb:EPJC77'2017} as
\begin{equation}\label{Massless (M.C)}
{\rm{Re}}~\Pi^L_n=\frac{g^2}{4\pi^2}\sum_{f} 
|q_{f}B|-\frac{g^2}{8\pi^2}\sum_{f}|q_{f}B|
~,\end{equation}
and for the physical quark masses, it becomes
\begin{equation}\label{Massive}
{\rm{Re}}~\Pi^L_n=-\frac{g^2}{4\pi^{2}T}\sum_{f}|q_{f}B|
\int^\infty_0dp_z
\frac{e^{\beta\sqrt{p^2_z+m^2_f}}}{\left(1+e^{\beta\sqrt{p^2_z
+m^2_f}}\right)^2}~.
\end{equation}
The medium contribution having quadratic dependence on 
the distribution function (\ref{G.S.E.D.D.}) does not yield 
any contribution to the real-part, i.e.
\begin{eqnarray}
{\rm{Re}}~\Pi^{\mu \nu}_{n^2}(k_{\parallel})=0~.
\end{eqnarray}

Thus the vacuum (\ref{Massless case}) and medium contributions 
(\ref{Massless (M.C)}) are combined together to give the longitudinal 
component due to the quark-loop in the limit of massless quarks
\begin{equation}
{\rm{Re}}~\Pi^L_{\rm {quark~loop}}=-\frac{g^2}{8\pi^2}\sum_{f} |q_{f}B|
~,\label{pi00}
\end{equation}
which depends on the magnetic field only in the 
SMF limit ($eB>>T^2$) and is independent of temperature even in the thermal 
medium.
The above form have also been calculated through the different
approaches~\cite{Fukushima:PRD93'2016,Bandyopadhyay:PRD94'2016,Mujeeb:EPJC77'2017}.

Similarly for the physical quark masses, the vacuum (\ref{Massive_case}) 
and medium contributions (\ref{Massive}) due to the quark loop are added to 
give the longitudinal component  in the static limit
\begin{equation}
{\rm{Re}}~\Pi^L_{\rm {quark~loop}}=-\frac{g^2}{4\pi^2T}
\sum_{f}|q_{f}B|\int^\infty_0dp_z
\frac{e^{\beta\sqrt{p^2_z+m^2_f}}}{\left(1+e^{\beta\sqrt{p^2_z
+m^2_f}}\right)^2}~,
\label{pi00m}
\end{equation}
which now depends on both magnetic field and temperature. However 
it becomes independent of temperature beyond 
a certain temperature~\cite{Mujeeb:EPJC77'2017}. 

We will now calculate the retarded/advanced gluon
self-energy tensor due to gluon loops using the $11$-component of matrix 
propagator for gluons
(\ref{temp_prop}) in a thermal medium. The longitudinal component of 
the same~\cite{Lata:PRD89'2014} 
is obtained by the HTL perturbation theory as
\begin{eqnarray}
\Pi^L_{\rm {gluon~loops}}(k)={g^{\prime}}^2 T^2 \left(\frac{k_{0}}{2\textbf{k}}\ln\frac
{k_{0}+\textbf{k} \pm i\epsilon}{k_{0}-\textbf{k} \pm i\epsilon}-1\right)~,
\label{self_energy_gluon}
\end{eqnarray}
with the prescriptions $+i\epsilon $ ($ -i\epsilon $) for the retarded
and advanced self-energies, respectively. Here we take $g^\prime=\sqrt{4
\pi \alpha_s^\prime (T)}$ as the 
one-loop strong running coupling, where the dominant scale
for gluonic degrees of freedom is the temperature so the 
renormalization scale is taken as $2 \pi T$.

Thus the real part of longitudinal component due to 
the gluon-loops in the static limit reduces to~\cite{Lata:PRD89'2014}
\begin{equation}\label{gluonloop}
{\rm{Re}}~\Pi^L_{\rm{ gluon~loops}}=-{g^{\prime}}^2 T^2
\end{equation}

Thus the Debye mass is obtained from static limit of quark-loop (\ref{pi00}) 
and gluon-loops (\ref{gluonloop}) contributions for the massless quarks 
\begin{equation}
m_{D}^2={g^{\prime}}^2 T^2+\frac{g^2}{8\pi^2} \sum_f |q_{f}B|~,
\label{debye_massless}
\end{equation}
Therefore the collective behaviour of the thermal 
medium in the presence of magnetic field is affected both by the temperature
and strong magnetic field, mainly through the gluon loop and quark loop
contributions, respectively. 
Similarly for the physical quark masses, the Debye mass is obtained 
\begin{eqnarray}
m_{D}^2={g^{\prime}}^2 T^2
+\frac{g^2}{4\pi^2T}\sum_f|q_fB|\int^\infty_0dp_z
\frac{e^{\beta\sqrt{p^2_z+m^2_f}}}{\left(1+e^{\beta\sqrt{p^2_z
+m^2_f}}\right)^2}~.
\label{debye_massive_1}
\end{eqnarray}

\begin{figure}[h]
\begin{center}
\includegraphics[width=6.5cm,height=6.5cm]{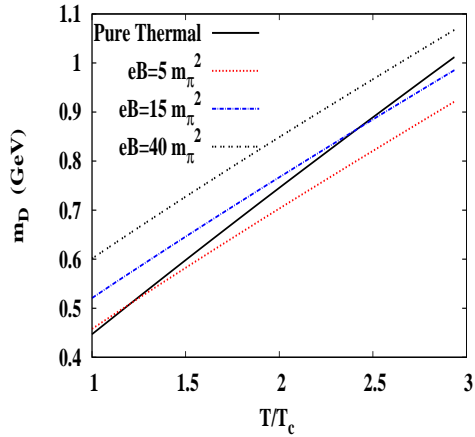}
\caption{Variation of Debye mass with temperature}
\end{center}
\end{figure}  
To see the competition between the temperature and the
magnetic field, we have plotted the Debye mass as a function 
of temperature at the different strength of magnetic fields in Figure 1. 
At lower temperatures the magnetic field contributes more to the 
screening mass than the temperature whereas as the temperature increases 
within the SMF limit ($eB \gg T^2$) the thermal
part plays more dominant role than the magnetic field unless the magnetic
field is sufficiently strong.

Therefore the real part of retarded gluon self-energy
(\ref{debye_massive_1}) 
gives the real-part of the retarded resummed gluon propagator
for realistic quark masses
\begin{eqnarray}
{\rm{Re}} D^L_R (k_0 \rightarrow 0)=\frac{1}{\textbf{k}^2+m_{D}^2}.
\label{resuumed_retarded}
\end{eqnarray}

\subsection{Imaginary part of retarded gluon self-energy}
Similar to the calculation of real part, the imaginary part of retarded 
self-energy is obtained from the real-time formalism
\begin{equation}
\rm{Im}~\Pi_R(k_0, {\bf k}) = \frac{\rm{Im} \bar\Pi(k_0, {\bf k})}{\varepsilon(k_0)},
\label{ret_re}
\end{equation}
where ${\rm{Im}}~\bar\Pi(k_0, {\bf k})$ is derived from the off-diagonal 
element of self-energy matrix as 
\begin{equation}
\rm{Im}~\bar\Pi(k_0, {\bf k}) =-\sinh(\beta k_0/2)\rm{Im}~\Pi_{12}(k_0, {\bf k}),
\label{re_12}
\end{equation}
and $\varepsilon(k_0)$ is the theta function. 
 
Like the evaluation of the real-part we will first 
evaluate the contribution due to quark-loop and then calculate 
for the gluon loops. Therefore the off-diagonal 
element (\ref{propagator_12}) of the propagator matrix (\ref{mag_prop}) for
quarks gives the $12$-component of self-energy matrix 
\begin{eqnarray}
i\Pi_{12}^L (k)  
&=&-\frac{g^2}{2}\sum_f\int\frac{dp_x dp_y}{(2\pi)^2}
e^{-\frac{(p+k)_\perp^2}{|q_fB|}}e^{-\frac{p_\perp^2}{|q_fB|}}\nonumber\\
&& \times \int dp_0 dp_z e^{\frac{\beta|p_0 + k_0|}{2}}
e^{\frac{\beta|p_0|}{2}} n_F(p_0) n_F(p_0 + k_0) L^{00}
\delta\left((p_\parallel + k_\parallel)^2 - m_f^2\right)
\delta(p_\parallel^2-m_f^2),
\label{pi1200}
\end{eqnarray}
wherein we use the equality $\sqrt{n_F(p_0)(1-n_F(p_0))}=e^{\frac{\beta p_0}{2}}
n_F(p_0)$ and the trace, $L^{00}$ is evaluated as
\begin{eqnarray}
L^{00} &=& 8\left[p_0(p+k)_0 + p_z(p+k)_z + m_f^2\right].
\label{trace}
\end{eqnarray}

The magnetic field 
again facilitates the calculation of the imaginary part 
by separating the momentum integration
into components perpendicular and parallel to the magnetic field, 
\begin{equation}
{\rm{Im}}~\Pi_{12}^L (k) = \frac{g^2}{2}\sum_f 
{\rm{Im}}~\Pi_\parallel(k_\parallel)~{\rm{Im}}~\Pi_\perp(k_\perp),
\label{pi_split}
\end{equation}
where the transverse component, $\Pi_\perp$ is integrated out as
\begin{equation}
{\rm{Im}}~\Pi_\perp (k_\perp) = 
\frac{|q_fB|}{8\pi}e^{-\frac{k_\perp^2}{2|q_fB|}},
\label{pi_perp}
\end{equation}
and after performing the $p_0$ integration, the parallel component is given by 
\begin{eqnarray}
{\rm{Im}}~\Pi_\parallel (k_\parallel) 
\label{pi_parallel}
&=& \int\frac{dp_z}{2\omega_p}e^{\frac{\beta|\omega_p|}{2}}
e^{\frac{\beta|k_0+\omega_p|}{2}}n(\omega_p)n(k_0 + \omega_p) L^{00}(p_0 
= \omega_p)\delta(k_0^2 - k_z^2 + 2p_0\omega_p - 2p_zk_z)\nonumber\\
&+& \int\frac{dp_z}{2\omega_p}e^{\frac{\beta|\omega_p|}{2}}
e^{\frac{\beta|k_0-\omega_p|}{2}}n(\omega_p)n(k_0 - \omega_p) L^{00}(p_0 
= -\omega_p)\delta(k_0^2 - k_z^2 - 2k_0\omega_p - 2p_zk_z).
\label{pi_pz}
\end{eqnarray}
Thus, in the static limit ($k_0 \rightarrow 0$), the longitudinal component of 
the imaginary part of retarded self-energy (\ref{ret_re}) assumes the form 
\begin{equation}
\lim_{k_0 \to 0}~\frac{\rm{Im} \Pi_{R}^L (\textbf{k})}{k_0} =-g^2 
\sum_f \frac{2 m_f^2}{T |k_z| E_{k_z/2}}
n_F(E_{k_z/2}) \left(1-n_F(E_{k_z/2}) \right) {\rm{Im}}~\Pi_\perp(k_\perp),
\label{impir_q00}
\end{equation}
with $E_{\frac{k_z}{2}} = \sqrt{m^2_f + k_z^2/4}$.

In weak coupling limit, the leading-order contribution 
in SMF limit comes from the momentum-transferred - 
$|\textbf{k}|^2\sim \alpha_s eB$, thus the exponential factor
in transverse component becomes unity -
$\exp{(-\frac{k^2_\bot}{2\mid q_fB \mid})}$ $\sim$ 1. Thus 
the transverse component of the imaginary part of the self-energy is
approximated into
\begin{equation}
{\rm {Im}}~\Pi_\perp \approx \frac{ \mid q_fB\mid}{8\pi}~, 
\label{perpendicular}
\end{equation}
and the dispersion relation is simplified too:
\begin{equation}
E_{\frac{k_z}{2}} \approx \frac{\mid k_z \mid}{2}.
\label{energy}
\end{equation}
Furthermore using the identity
\begin{equation}
n_F(E_{\frac{k_z}{2}})\Big[1-n_F(E_{\frac{k_z}{2}})\Big]
=\frac{1}{2\big[1+\cosh(\beta E_{\frac{k_z}{2}}\big]},
\end{equation}
the imaginary component is rewritten as
\begin{equation}
\lim_{k_0 \to 0}~\left[ \frac{\rm{Im}~\Pi_{R}^L (\textbf{k})}{\rm k_0} 
\right]=
-\frac{g^2}{4\pi T} \sum_f m^2_f \mid q_fB\mid
\frac{1}{k_z^2\big[1+\cosh(\beta E_{\frac{k_z}{2}}\big]},
\label{self_retarded4}
\end{equation}

Moreover in SMF limit the longitudinal component ($\mid k_z \mid$)
of the momentum is of the order $(\alpha_s eB)^{1/2}$, which is much smaller
than the temperature ($<<T$). Therefore, the imaginary
component of retarded self energy due to quark-loop takes 
further lucid form
\begin{equation}
\lim_{k_0 \to 0}\Big[\frac{\rm{Im}\Pi^L_R(\textbf{k})}{k_0}\Big]_{\rm{quark~loop}}=
-g^2\frac{\sum_f m^2_f\mid q_fB\mid}{8\pi T}\frac{1}{k^2_z}~, 
\label{self_retarded1}
\end{equation}

Similarly we will now calculate the imaginary part due to the gluon loops 
from the off-diagonal element of the self-energy matrix by the
off-diagonal element of gluon propagator matrix (\ref{temp_prop}). However,
it will be easier to calculate it directly from the imaginary part of the retarded self-energy 
from the gluon-loop contribution (\ref{self_energy_gluon}). Thus using
the identity
\begin{equation}\label{identity}
\frac{1}{x\pm{y}\pm{i\epsilon}}={\rm{P}}\left(\frac{1}
{x\pm{y}}\right)\mp{i\pi{\delta(x\pm{y})}}
~,\end{equation}
the imaginary part due to the gluon-loop is extracted from
(\ref{self_energy_gluon})
\begin{eqnarray}
\lim_{k_0 \to 0}\Big[\frac{\rm{Im}\Pi^L_R(\textbf{k})}
{k_0}\Big]_{\rm{gluon~loops}}=-{g^{\prime}}^2\frac{\pi T^2}{2}\frac{1}{\textbf{k}}.
\end{eqnarray}

Thus the longitudinal component of the imaginary part of 
gluon self-energy due to both quark and gluon-loop always
factorizes into $k_0$ times $\rm{Im}\Pi^L_R (\bf k)$ so 
it vanishes in the static limit ($k_0 \to 0$). Therefore, using
the factors in (\ref{factor}, \ref{retarded_advanced}), the resummed symmetric 
propagator (\ref{lon_sym}) in the static limit reduces to
\begin{eqnarray}
D^{L}_{S}(\textbf{k})&=&[1+2n_{B}(k_0)]~{\rm{sgn}}(k_0) \left(D^{L}_{R}(k)-D^{L}_{A}(k)
\right) \nonumber\\
&=& i 4T \frac{{\rm{Im}} \Pi^{L}_{R}(\textbf{k})}{\left[\textbf{k}^2-\rm{Re} 
\Pi^{L}_{R}(\textbf{k})\right]^2},
\end{eqnarray}
which is however decomposed into the contributions due to the quark and 
gluon loop
\begin{eqnarray}
D^L_S(\textbf{k})&=& D^L_S(\textbf{k})_{\rm{quark ~loop}}+
D^L_S(\textbf{k})_{\rm{gluon~loop}}
\end{eqnarray}
with
\begin{eqnarray}
D^L_S(\textbf{k})_{\rm{quark ~loop}} &=& -
\frac{ig^2}{2\pi k^2_z}\frac{\sum_f 
\mid q_fB\mid m^2_f}{({\textbf{k}}^2+m^2_D)^2}
\label{resummed_symmetric_quark}\\
D^L_S(\textbf{k})_{\rm{gluon~loop}} &=& \frac{-2i\pi  {g^{\prime}}^2 T^3}{\textbf{k}({\textbf{k}}^2+m_D^2)^2}~.
\label{resummed_symmetric_gluon}
\end{eqnarray}

\section{Heavy quark potential}
The derivation of potential between a heavy quark $Q$ and its anti-quark
($\bar Q$) from effective field theory, {\em namely} pNRQCD 
may not be plausible because the hierarchy of non relativistic 
scales and thermal scales assumed in weak coupling EFT calculations may not 
be satisfied. Even in the first principle QCD study, the 
adequate quality of the data is not available in the present lattice
correlator studies so one may use the potential model to circumvent the 
problems.
Since the mass of the heavy quark ($m_Q$) is very large, so the 
requirement - $m_Q \gg T \gg \Lambda_{QCD}$ 
is satisfied for the description of the interactions between a pair of heavy 
quark and anti-quark at finite temperature in strong magnetic 
field limit in terms of quantum mechanical 
potential. Thus we can obtain the medium-modification to the vacuum potential
in the presence of magnetic field by correcting both its short and 
long-distance part with a dielectric function $\epsilon (\textbf{k})$ as 
 \begin{equation}
V(r;T,B)=\int\frac{d^3\textbf{k}}{(2\pi)^{3/2}}
({e^{i\textbf{k}.\textbf{r}}-1})\frac{V(\textbf{k})}{\epsilon(\textbf{k})},
\label{pot_defn}
\end{equation}
where we have subtracted a $r$-independent term to renormalize the heavy
quark free energy, which is the perturbative free energy of quarkonium at
infinite separation. The Fourier transform,  $V(\textbf{k})$  of the 
Cornell potential is given by
\begin{equation}
{V}(\textbf{k})=-\frac{4}{3}\sqrt{\frac{2}{\pi}} \frac{\alpha_s}{\textbf{k}^2}-\frac{4\sigma}
{\sqrt{2 \pi} \textbf{k}^4},
\label{ft_pot}
\end{equation}
and the dielectric permittivity, $\epsilon(\mathbf k)$ 
encodes the effects of deconfined medium in the presence of magnetic 
field, which is going to be calculated next.
\subsection{The complex permittivity for a hot QCD medium in a strong magnetic
field}
The dielectric permittivity is defined by the static limit of 11-component 
of longitudinal resummed gluon propagator by the following equation
\begin{equation}
\frac{1}{\epsilon (\bf{k})}=\displaystyle
{\lim_{k_0 \rightarrow 0}}{\textbf{k}}^{2}D_{11}^{L}(k_{0},
\textbf{k}),
\label{dielectric}
\end{equation}
where the real and imaginary parts of $D^{L}_{11}(\textbf{k})$ are obtained
by the retarded (or advanced) and symmetric propagator, respectively
\begin{eqnarray}
\rm{Re} D^{L}_{11}(\textbf{k})&=&\rm{Re} D^{L}_{R} (\textbf{k}) \nonumber\\
\rm{Im} D^{L}_{11}(\textbf{k})&=&\rm{Im} \frac{D^{L}_{S}}{2}
(\textbf{k})  ,
\label{imaginary_propagator}
\end{eqnarray}
which will in turn gives the real and imaginary part of dielectric 
permittivity, respectively.

Thus the static limit of resummed retarded propagator (\ref{resuumed_retarded}) 
gives the real part of dielectric permittivity 
\begin{equation}
\frac{1}{{\rm Re}~\epsilon(\bf{k})}=\frac{\textbf{k}^2}{\textbf{k}^2+m_{D}^2}.
\label{real_dielectric}
\end{equation}
Similarly the static limit of resummed symmetric propagators 
(\ref{resummed_symmetric_quark},~\ref{resummed_symmetric_gluon}) gives 
the imaginary part of dielectric permittivity, due to quark and 
gluon loop contributions
\begin{eqnarray}
\frac{1}{\rm{Im}~{\epsilon (\bf{k})}_{\rm quark~loop}}&=&-\frac{g^2}{4\pi} 
\sum_f m_f^2 \mid q_fB\mid 
\frac{{\bf k}^2}{k^2_z({\bf{k}}^2+m_D^2)^2}~, \label{img_dielectric_quark}\\
\frac{1}{{{\rm{Im}}~\epsilon (\bf k)}_{\rm gluon~loop}} &=&-{g^{\prime}}^2\pi T^3 
\frac{{\bf{k}}^2}{\bf{k}({{\bf{k}}^2+m_D^2)}^2}~,
\label{img_dielectric_gluon}
\end{eqnarray}
respectively. 

Therefore the real and imaginary part of dielectric permittivities
give the real and imaginary part of 
the complex potential, respectively in the next subsection.

\subsection{Real and Imaginary Part of the potential}
The real-part of dielectric permittivity (\ref{real_dielectric}) 
is substituted into the definition (\ref{pot_defn}) to give the real part of 
$Q \bar Q$ potential in the presence of strong magnetic field~\cite{Mujeeb:EPJC77'2017} 
(with $\hat{r}=rm_{D}$)
\begin{eqnarray}
\rm{Re} V(r;T,B)&=&-\frac{4}{3}\alpha_s m_{D} \frac{e^{-\hat{r}}}{\hat{r}} 
+\frac{2\sigma}{m_{D}} \frac{(e^{-\hat{r}}-1)}{\hat{r}} \nonumber\\
&-&\frac{4}{3}\alpha_s m_{D}+\frac{2\sigma}{m_{D}}~,
\label{real_potential}
\end{eqnarray}
where the dependence of temperature and magnetic field enter through the 
Debye mass. The nonlocal terms insure the potential in medium $V(r;T,B)$ 
to reduce to the potential in $(T, B) \rightarrow 0 $ limit, which are, 
however, required to compute the masses of quarkonium states.
The additional effect due to the strong magnetic field 
on the potential in a hot QCD medium is displayed as a function 
of interparticle distance ($r$) for different strength of magnetic fields 
in Figure 2, after excluding the constant terms from (\ref{real_potential}).
The solid line represents the potential in a pure thermal medium ({\em 
i.e.} in the absence of magnetic field) whereas the dashed 
and dotted lines denote the effect of strong magnetic 
fields, 10 and 25 $m_\pi^2$ to a thermal medium, respectively. We 
have found that the magnetic field  (eB=10 $m_\pi^2$)
affects the linear string term more than the Coulomb term, as a result, the 
overall potential at small and intermediate $r$ becomes less screened than 
the potential in pure thermal medium. However, further increase of magnetic 
field ({\em i.e.} eB= 25 $m_\pi^2$), the 
potential becomes less attractive than eB=10 $m_\pi^2$.
However for large $r$ the effect of magnetic field diminishes
gradually.

\begin{figure}[h]
\begin{center}
\begin{tabular}{c c}
\includegraphics[width=6.5cm,height=6.5cm]{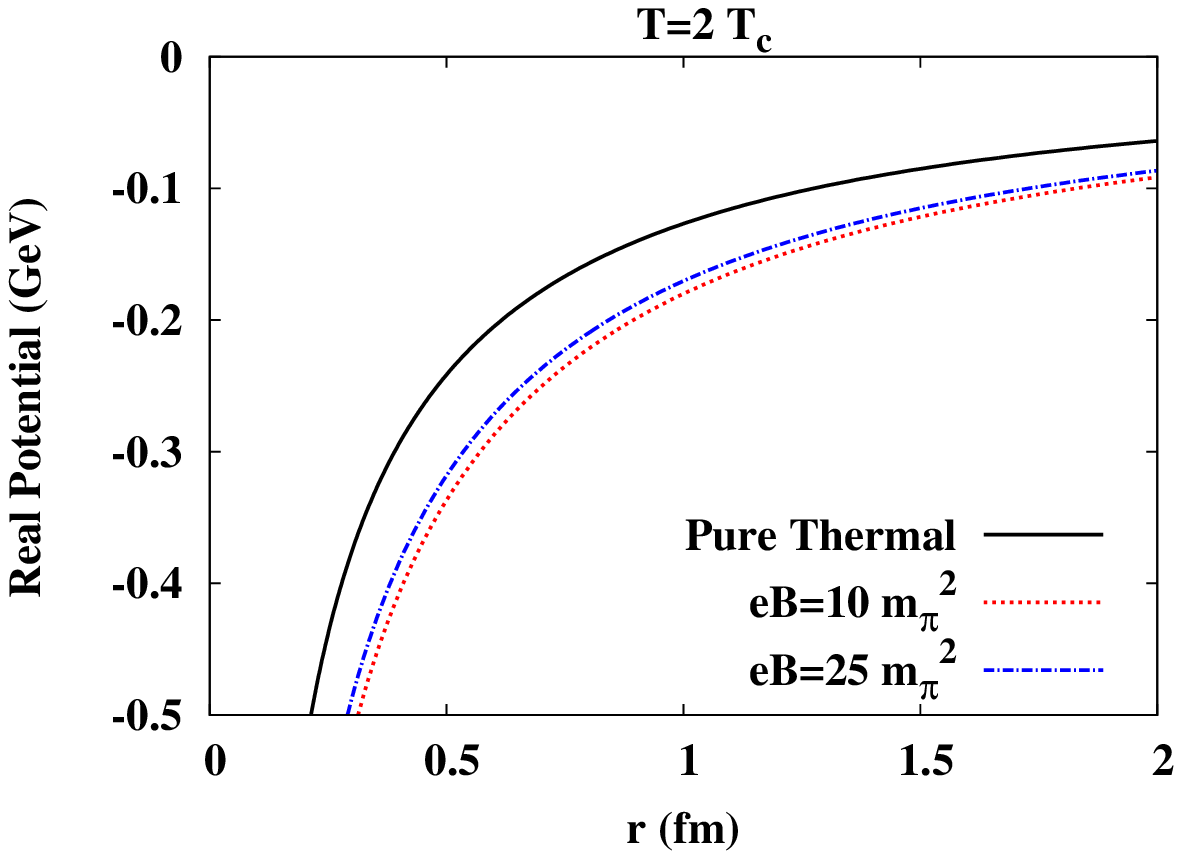}&
\includegraphics[width=6.5cm,height=6.5cm]{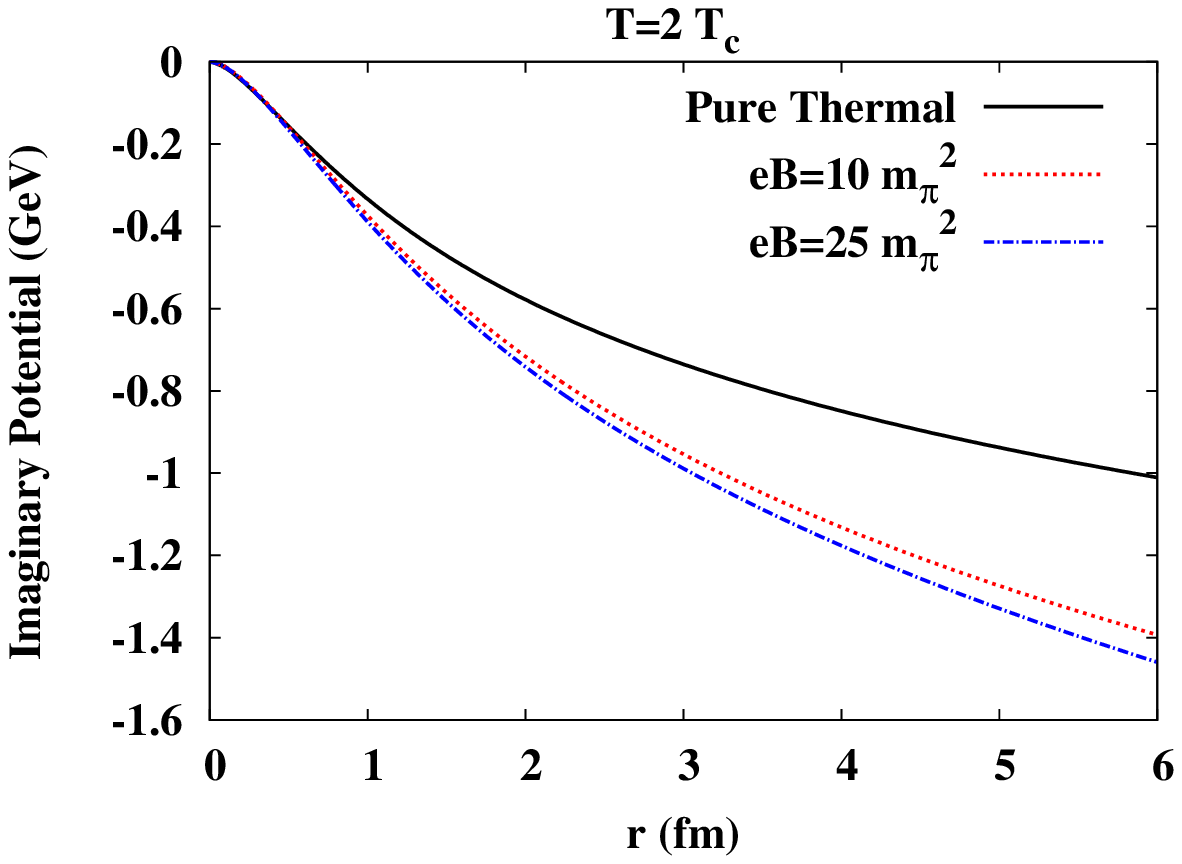}\\
\end{tabular}
\caption{Real (left) and imaginary (right) part of the potential}
\end{center}
\label{fig1}
\end{figure}
Similarly the imaginary part of the potential is obtained
by plugging the imaginary part of dielectric permittivities due to quark-loop 
(\ref{img_dielectric_quark}) and gluon-loop (\ref{img_dielectric_gluon}) 
contributions into the definition of potential (\ref{pot_defn}). The imaginary 
component of the potential consists of Coulomb and string terms
\begin{equation}
\rm{Im}~V(r;T,B) =\rm{Im}~V_C(r;T,B)+\rm{Im}~V_S(r;T,B), 
\end{equation}
where each term is again split into quark-loop ($q$) and gluon-loop ($g$) 
contributions. We will first calculate due to the quark loop 
from (\ref{img_dielectric_quark})
\begin{eqnarray}
\rm{Im} V_C^q (r;T,B)&=& \int \frac{d^3\textbf{k}}{(2\pi)^{3/2}}
\left(e^{ik.r}-1\right)\left(-\frac{4}{3}\sqrt{\frac{2}{\pi}} 
\frac{\alpha_s}{{\textbf{k}}^2}\right)\left(-\frac
{g^2{\textbf{k}}^2}
{4\pi k^2_z}\frac{\sum_f 
\mid q_fB\mid m^2_f}{{({\textbf{k}}^2+m^2_D)}^2}
\right)\nonumber\\
&=& \frac{\alpha_sg^2}{3\pi^2}
\left(\sum_f \mid q_fB\mid m^2_f\right)
I_C,
\label{alpha_quark}
\end{eqnarray}
where the momentum integral, $I_C$ is integrated as 
\begin{eqnarray}
I_C&=&\int_0^{\infty} \frac{dk}
{{({\textbf{k}}^2+m^2_D)}^2} \int_{-1}^{1} 
dx\frac{(e^{ikrx}-1)}{x^2}\nonumber\\
&=&\int_0^{\infty} \frac{dk}
{{({\textbf{k}}^2+m^2_D)}^2}\left[2-2\cos(kr)-2kr~Si(kr)\right]\nonumber\\
&\equiv&I_{C1}+I_{C2}+I_{C3},
\end{eqnarray}
where 
\begin{eqnarray}
I_{C1}&=&2 \int_0^{\infty} \frac{dk}{{({\textbf{k}}^2+m^2_D)}^2}
=\frac{\pi}{2m_D^3}\\
I_{C2}&=&-2 \int_0^{\infty} \frac{\cos kr~dk} {{({\textbf{k}}^2+m^2_D)}^2} 
 =- \left[\frac{\pi e^{-\hat{r}}}{2m_D^3}+
\frac{\hat{r}\pi e^{-\hat{r}}}{2m_D^3} \right] \\
I_{C3}&=&-2r \int_0^{\infty} \frac{dk~k}{{({\textbf{k}}^2+m^2_D)}^2} Si(kr) \nonumber\\
&=&-2\frac{\hat{r}}{m_D} \int_0^{\infty} \frac{dk~k}{{({\textbf{k}}^2+m^2_D)}^2} 
\int_0^{kr} dx \frac{\sin x}{x}~, 
\end{eqnarray}
respectively.
Similarly the string part of the imaginary potential is 
\begin{eqnarray}
\rm{Im} V_S^q (r;T,B) &=& \int \frac{d^3\textbf{k}}{(2\pi)^{3/2}}
\left(e^{ik.r}-1\right)\left(-\frac{4\sigma}
{\sqrt{2\pi}{\textbf{k}}^4} 
\right)\left(-\frac
{g^2{\textbf{k}}^2}
{4\pi k^2_z}\frac{\sum_f 
\mid q_fB\mid m^2_f}{{({\textbf{k}}^2+m^2_D)}^2}
\right)\nonumber\\
&=& \frac{\sigma g^2}{2\pi^2}
\left(\sum_f \mid q_fB\mid m^2_f\right)
I_S,
\label{sigma_quark}
\end{eqnarray}
where the integral, $I_S$ is evaluated as
\begin{eqnarray}
I_S &=&\int_0^{\infty} \frac{dk}
{{\textbf{k}}^2{({\textbf{k}}^2+m^2_D)}^2}
\int_{-1}^{1} dx\frac{(e^{ikrx}-1)}{x^2} \nonumber\\
&=&\int_0^{\infty} \frac{dk}
{{\textbf{k}}^2{({\textbf{k}}^2+m^2_D)}^2}\left[2-2\cos(kr)-2krSi(kr)\right]\nonumber\\
&\equiv & I_{S1} +I_{S2},
\label{integration_sigma}
\end{eqnarray}
where $I_{S1}$ and $I_{S2}$ are given by
\begin{eqnarray}
I_{S1}&=&\int_0^{\infty} \frac{dk}
{{\textbf{k}^2{({\textbf{k}}^2+m^2_D)}^2}}
\left( 2-2cos(kr) \right)\nonumber\\
&=&\frac{\pi}{2m_D^5}\left[\hat{r}e^{-\hat{r}}-3(1-e^{-\hat{r}})
+2\hat{r}\right]\\
I_{S2} &=& -2\frac{\hat{r}}{m_D} \int_0^{\infty} \frac{dk}
{{{k}{({\textbf{k}}^2+m^2_D)}^2}} \int_0^{kr} \frac{\sin x}{x} dx,
\end{eqnarray}

Next we calculate the imaginary part due to the gluon loop 
contribution (\ref{img_dielectric_gluon}) for the Coulomb and
string terms, respectively~\cite{Lata:PRD89'2014} as
\begin{eqnarray}
\rm{Im} V^{g}_C( r;T,B)&=&-\frac{8{\alpha_s}^{\prime} T}{3}  \int_0^{\infty} \frac{dz~z}{(z^2+1)^2}
\left(1-\frac {\sin{z\hat r}}{z\hat r}\right)\label{alpha_gluon}\\
\rm{Im} V_S^{g}(r;T,B) &=&-\frac{4\sigma T }{m_D^2}\int_0^{\infty}
\frac{dz}{z(z^2+1)^2}
\left(1-\frac {\sin{z\hat r}}{z\hat r}\right)~,
\label{sigma_gluon}
\end{eqnarray}
where the Debye mass is given by Eq.(\ref{debye_massive_1}). 

Thus the equations (\ref{alpha_quark}) and (\ref{alpha_gluon}) give the
Coulombic contribution whereas the equations (\ref{sigma_quark}), 
(\ref{sigma_gluon}) give the string contribution 
\begin{eqnarray}
{\rm{Im}}~V_C(r;T,B)&=& \rm{Im} V^{q}_C( r;T,B)+\rm{Im} V^{g}_C( r;T,B)\\
{\rm{Im}}~V_S(r;T,B)&=& \rm{Im} V^{q}_S( r;T,B)+\rm{Im} V^{g}_S( r;T,B)
\end{eqnarray}
to the imaginary component of the potential, respectively. Like the real-part 
of potential, how does the imaginary part get affected by the additional 
presence of 
magnetic field we have plotted it as a function of interquark distance 
in the right panel of Figure 2. In pure thermal medium (denoted by solid line), both Coulomb and string 
term are larger powers of $\hat{r}$ and counter each other, resulting the 
overall magnitude very small. Now the strong magnetic 
field not only
reduces the power of $\hat{r}$ in both terms compared to the 
pure thermal medium, it induces Coulomb and string
terms to contribute additively, resulting the overall magnitude 
of imaginary part larger. The above observation ultimately translates into 
the enhancement of thermal width of resonance states due to the ambient 
strong magnetic field.

\section{Properties of Quarkonia}
\subsection{Wavefunction and Binding Energy}
To investigate the properties of quarkonia in a strong magnetic field, 
we first solve the Schr\"odinger equation numerically by employing the 
real part of the potential (\ref{real_potential}) 
to see how the eigenstates to $J/\psi$, 
$\psi^\prime$ and $\chi_c$ states in a thermal QCD medium get affected by 
the presence of strong magnetic field in figures 3-5, respectively. 
In the presence of magnetic
field both wavefunction and probability distribution of quarkonia becomes
sharply peaked compared to quarkonia in absence of magnetic field.
\begin{figure}[h]
\begin{center}
\begin{tabular}{c c}
\includegraphics[width=6.5cm,height=6.5cm]{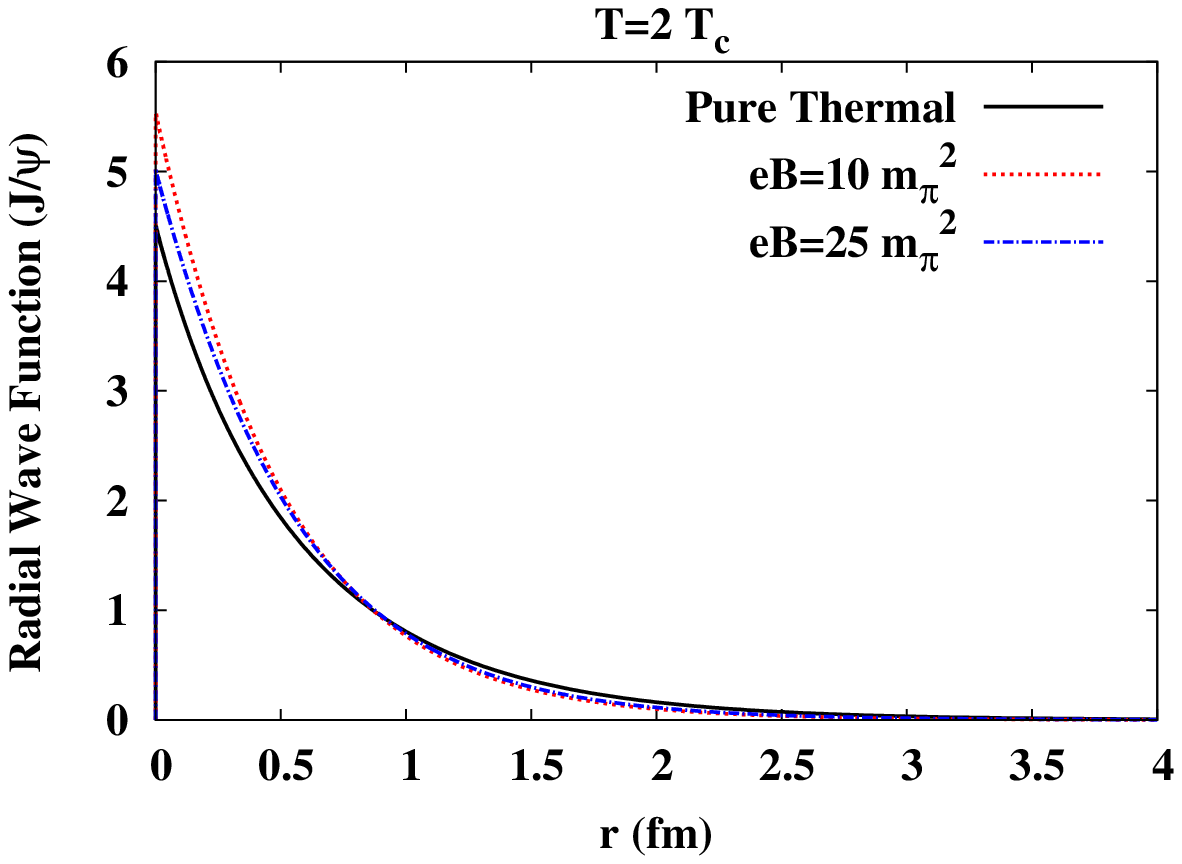}&
\includegraphics[width=6.5cm,height=6.5cm]{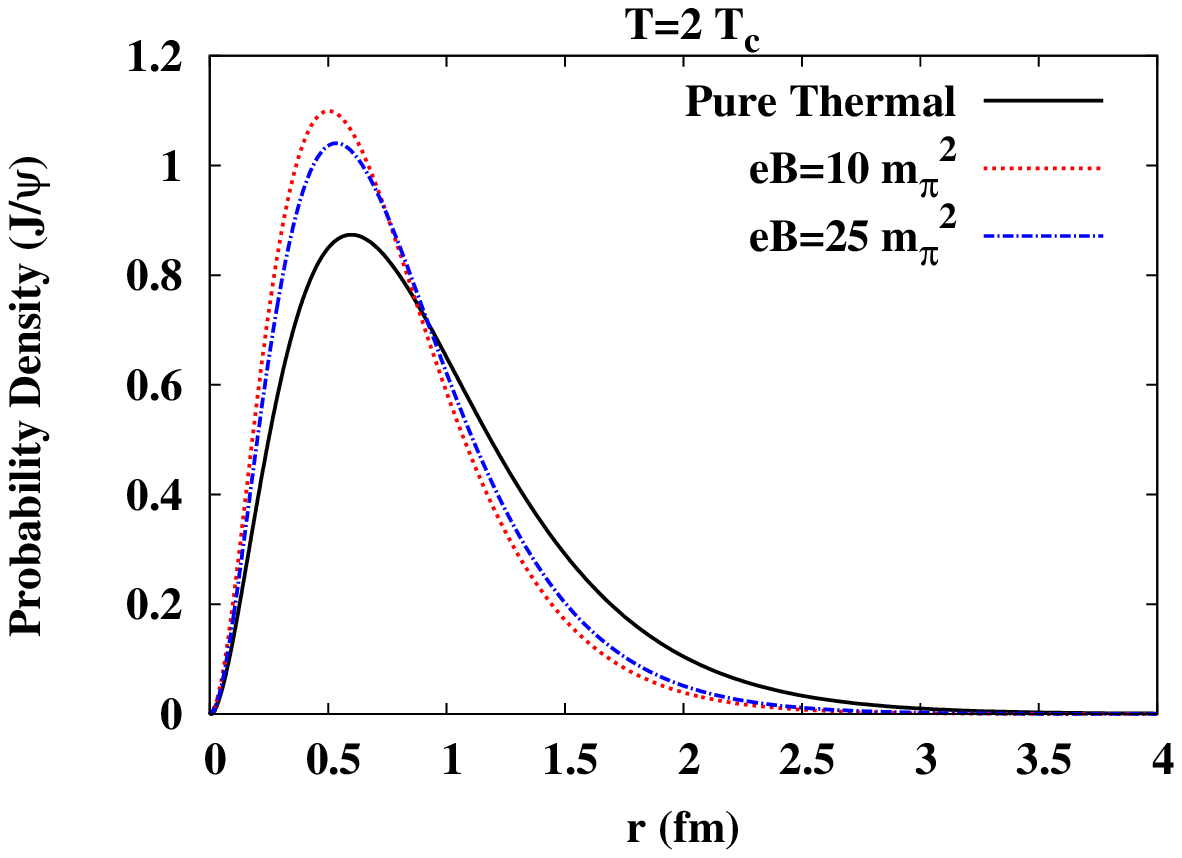}\\
\end{tabular}
\caption{The wavefunction and the radial probability density of $J/\psi$ state}
\end{center}
\label{fig1}
\end{figure}

\begin{figure}[h]
\begin{center}
\begin{tabular}{c c}
\includegraphics[width=6.5cm,height=6.5cm]{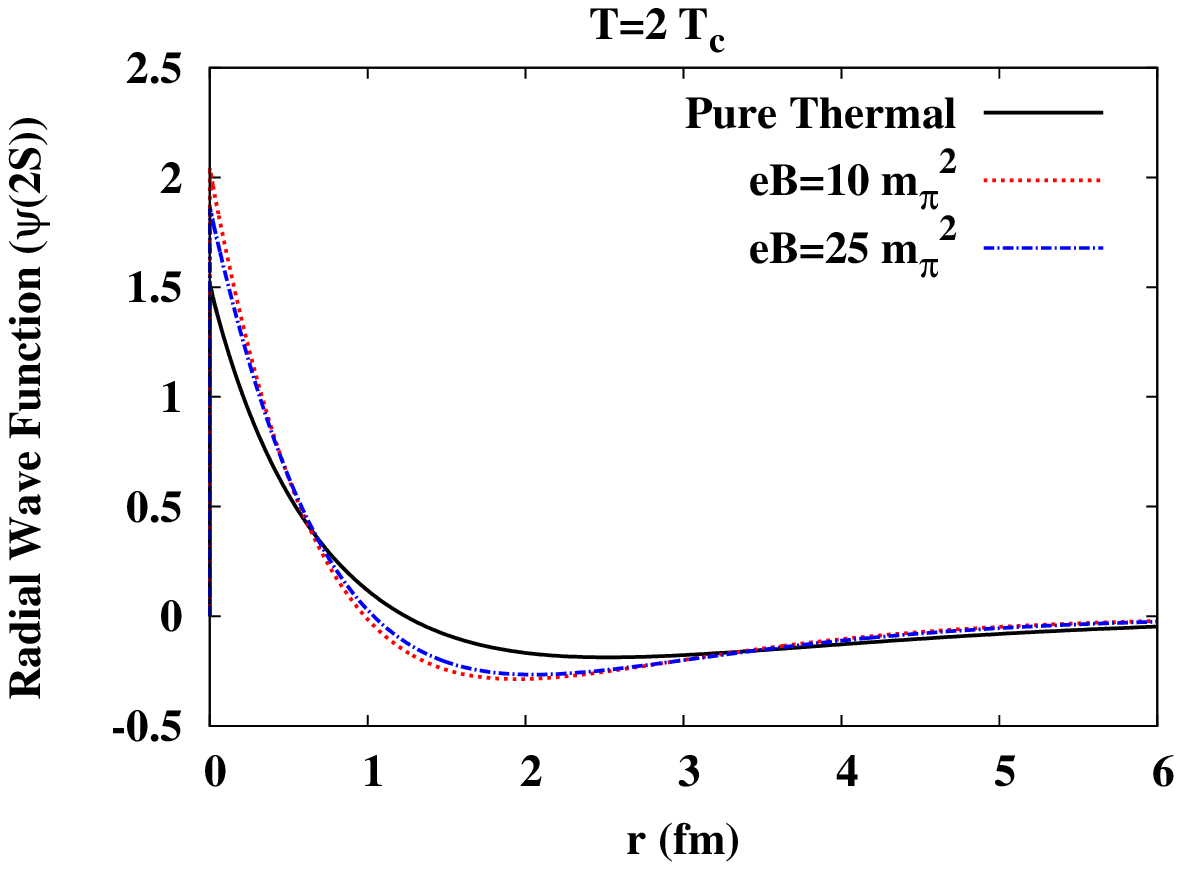}&
\includegraphics[width=6.5cm,height=6.5cm]{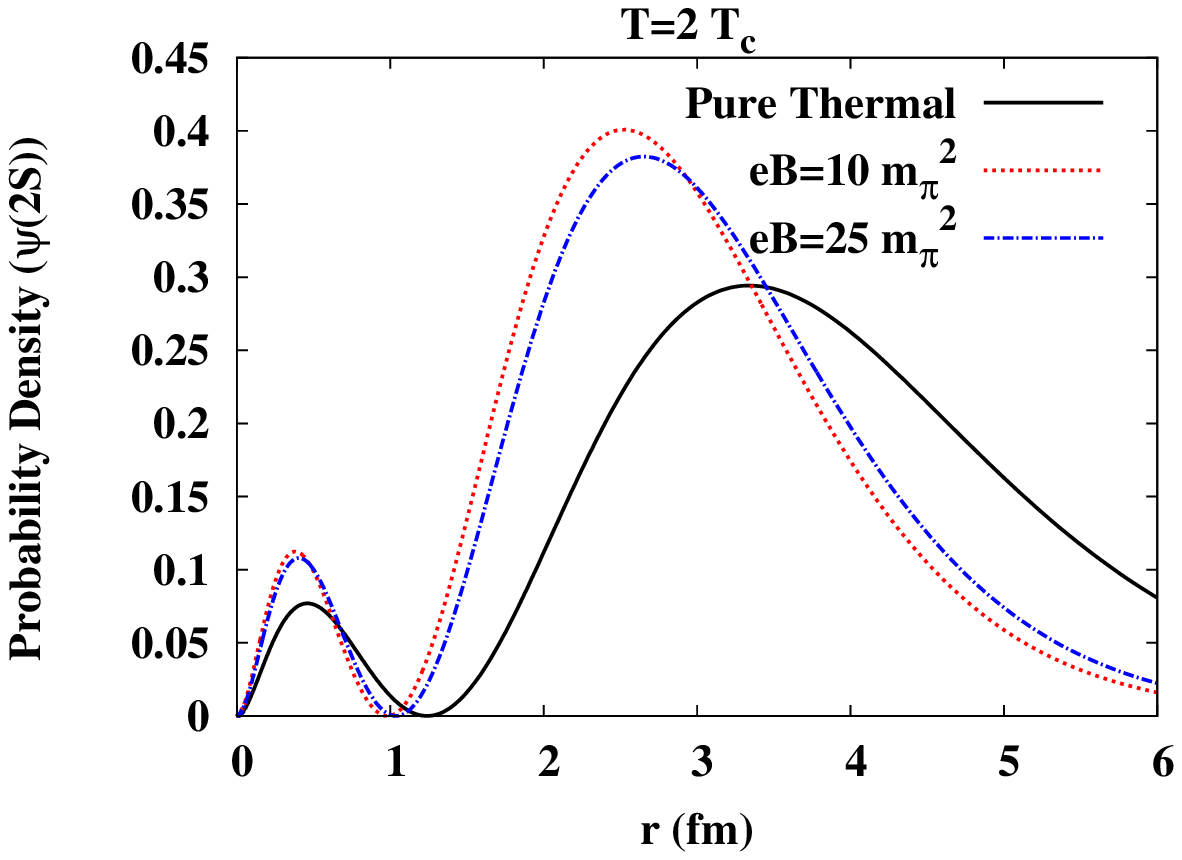}\\
\end{tabular}
\caption{The wavefunction and the radial probability density of $\psi^\prime$ 
state}
\end{center}
\label{fig1}
\end{figure}   

\begin{figure}[h]
\begin{center}
\begin{tabular}{c c}
\includegraphics[width=6.5cm,height=6.5cm]{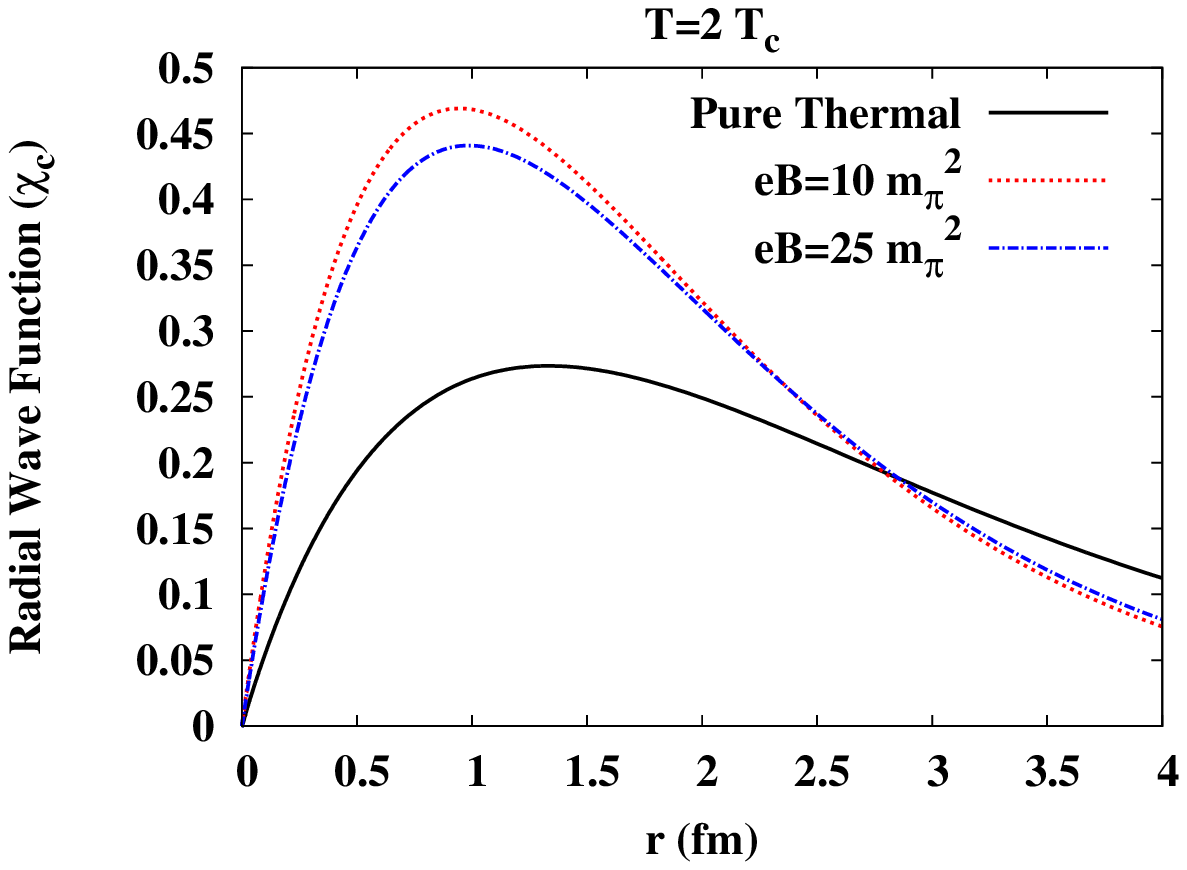}&
\includegraphics[width=6.5cm,height=6.5cm]{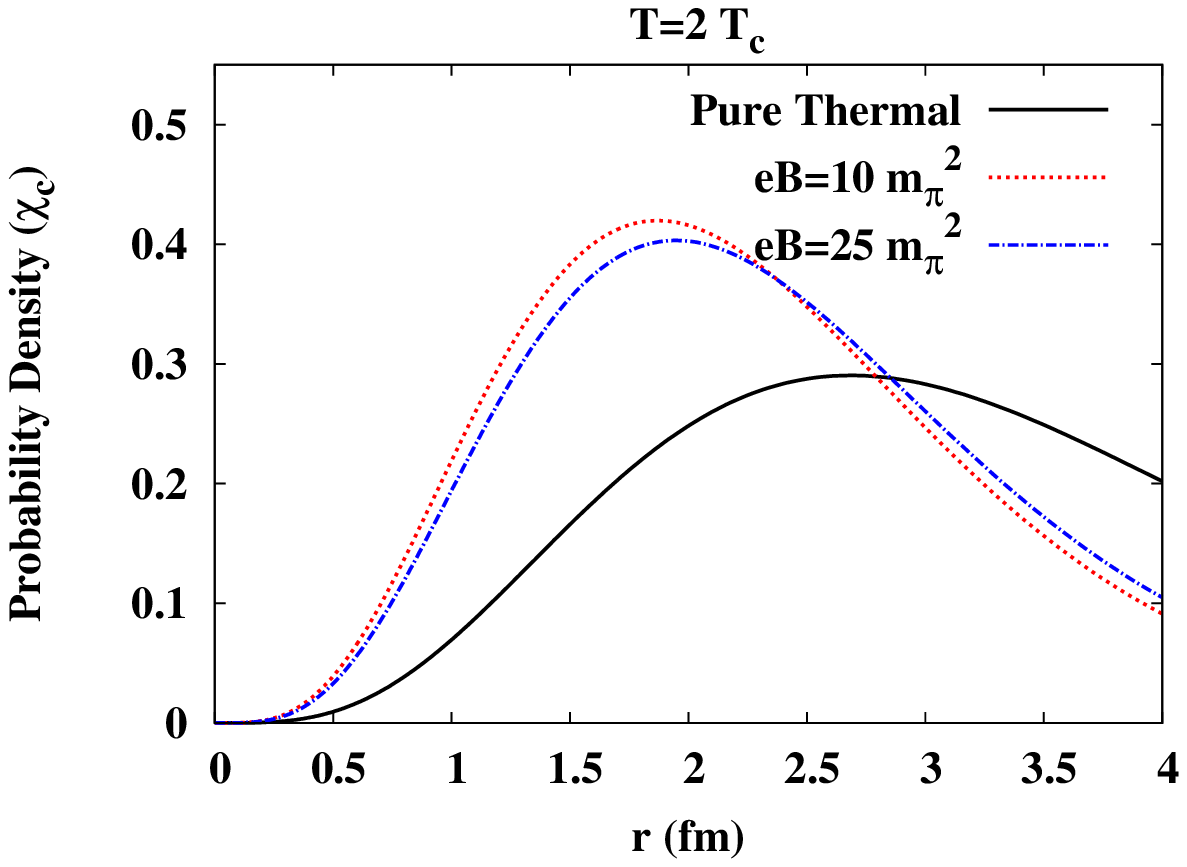}\\
\end{tabular}
\caption{The wavefunction and the radial probability density of $\chi_c$ state}
\end{center}
\label{fig1}
\end{figure}  
Thus the medium effects encoded into the wavefunctions ($\Phi(r)$) and the 
corresponding probability distributions explore how the average 
size of a particular quarkonia ($\sqrt{{r_i}^2}$ =${(\int d \tau ~r^2~{\mid 
\Phi_i(r) \mid}^2)}^{1/2}$) get affected due to a thermal medium 
in absence (presence) of magnetic field in left (right) panel of Figure 6, 
respectively. The magnetic field in general causes swelling of all resonances 
unless the temperature is very large (Figure 7).
\begin{figure}[h]
\begin{center}
\begin{tabular}{c c}
\includegraphics[width=6.5cm,height=6.5cm]{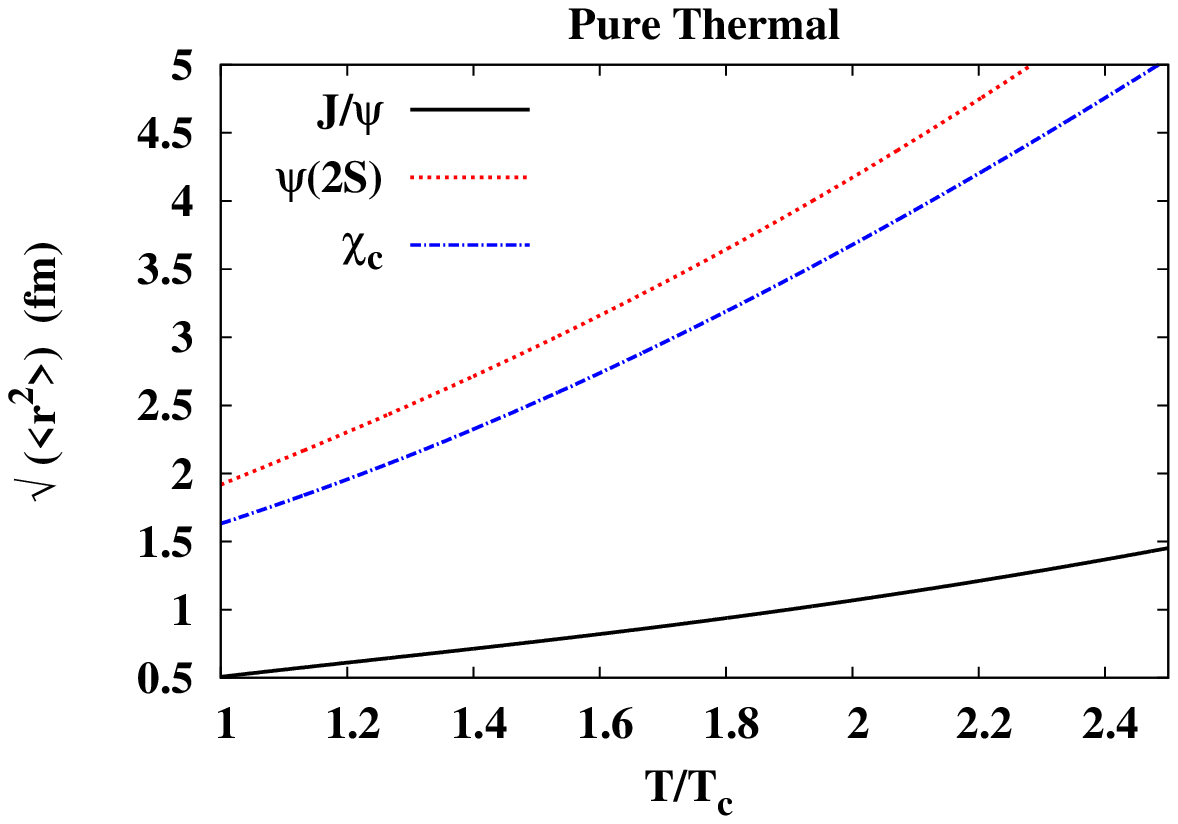}&
\includegraphics[width=6.5cm,height=6.5cm]{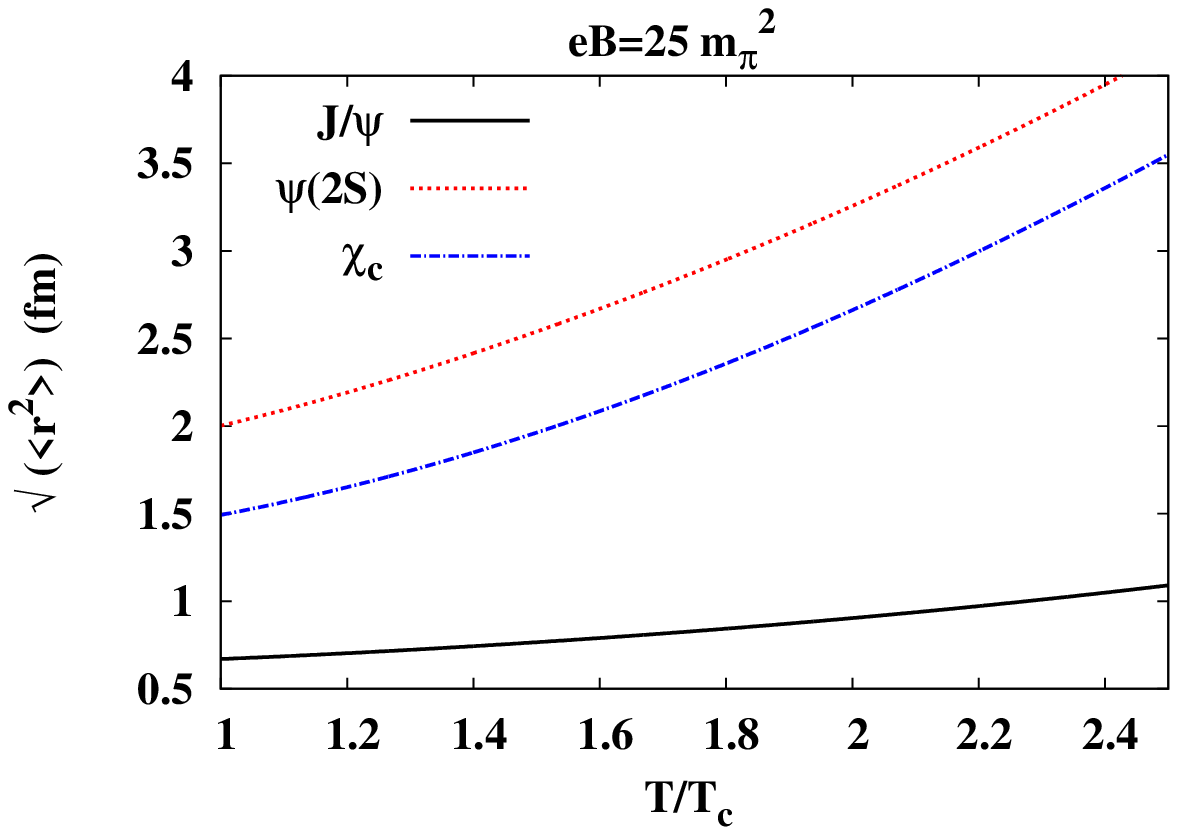}\\
\end{tabular}
\caption{The average size ($\sqrt{r^2}$) of quarkonia in pure thermal 
medium (left) and then thermal medium in presence of strong magnetic 
field (right)}
\end{center}
\label{fig1}
\end{figure}   
\begin{figure}[h]
\begin{center}
\includegraphics[width=6.5cm,height=6.5cm]{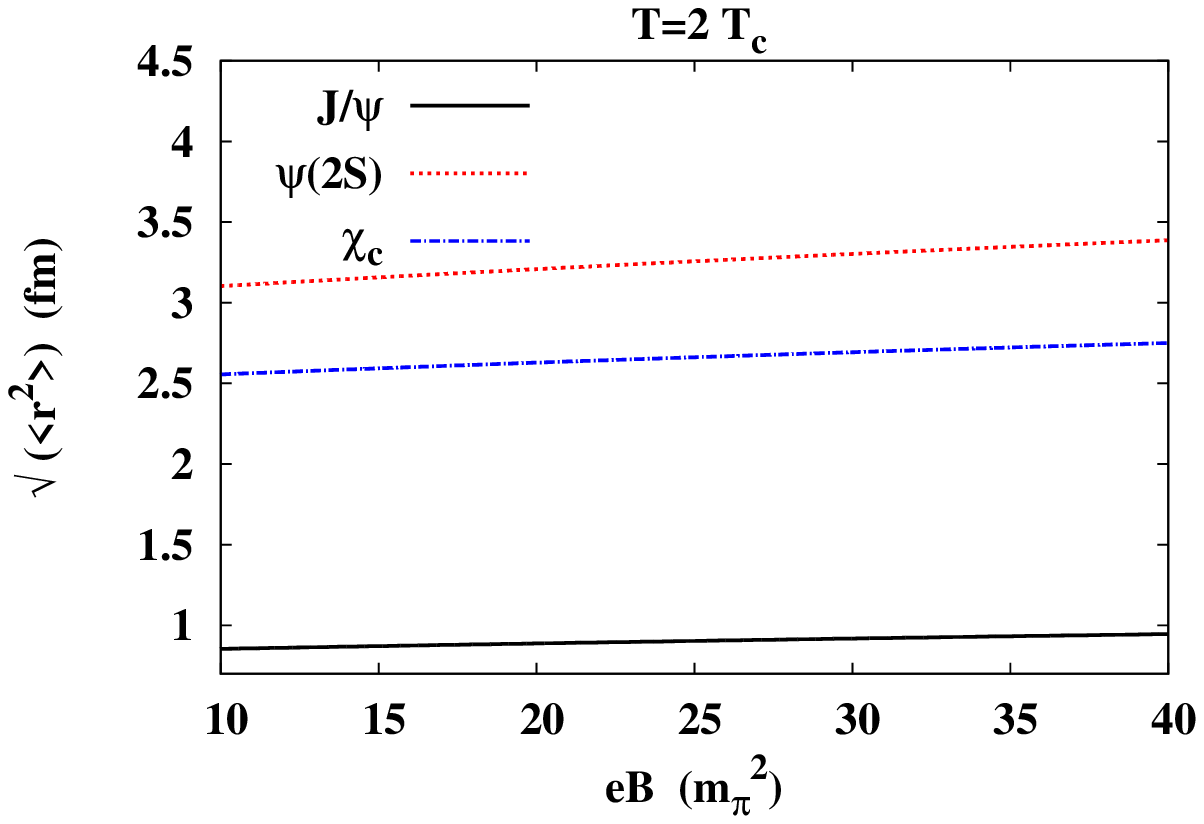}
\caption{Variation of the size of quarkonia with the magnetic field 
at a fixed temperature.}
\end{center}
\label{fig1}
\end{figure}   

Finally we have studied how the binding energies of
quarkonia change with the temperature in absence (presence)
of magnetic field in left (right) panel of Figure 8, respectively.
The immediate observation is that the magnetic field causes the
binding energy to decrease with the temperature slowly, compared to the
medium in absence of magnetic field.
Moreover the competition between the scales associated to the temperature and 
magnetic field affects the binding of quarkonia discriminately, {\em viz.}
$J/\psi$ becomes less bound and $\chi_c$ becomes more bound
due to the presence of magnetic field. However, the binding energy
decreases with the magnetic field too (Figure 9).

\begin{figure}[h]
\begin{center}
\begin{tabular}{c c}
\includegraphics[width=6.5cm,height=6.5cm]{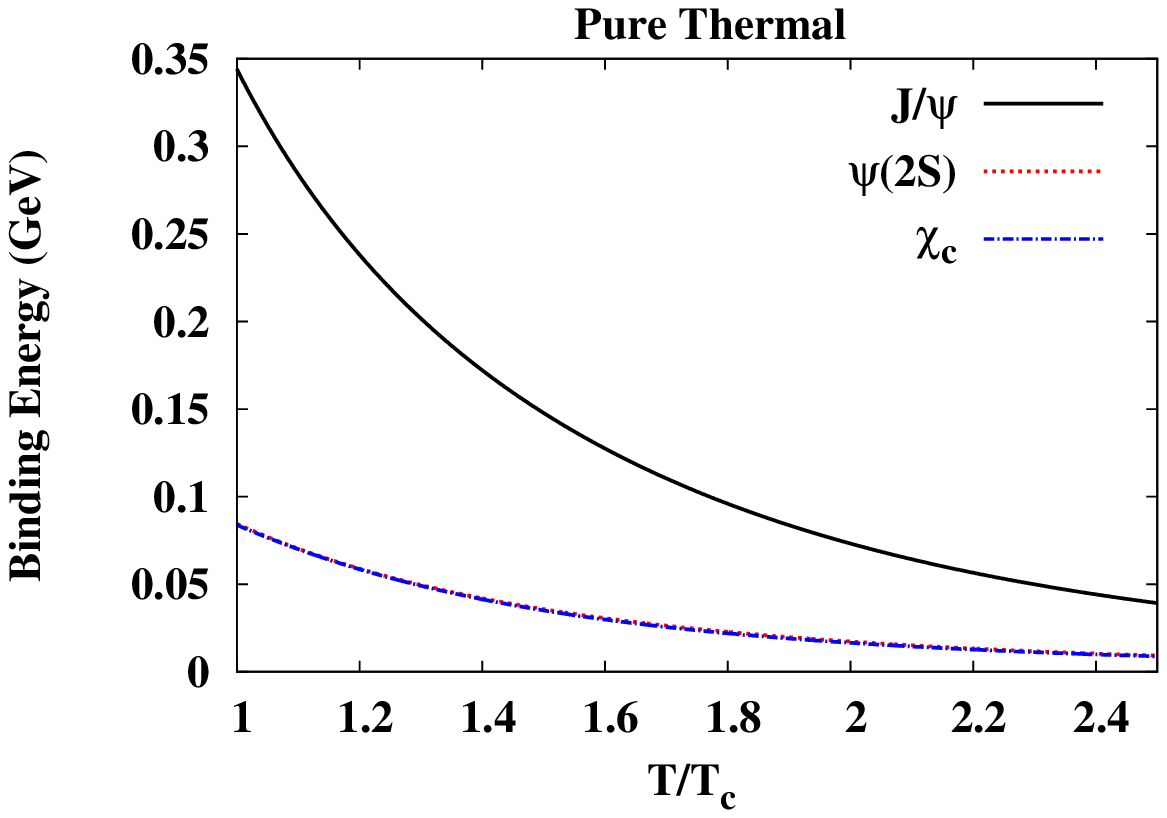}&
\includegraphics[width=6.5cm,height=6.5cm]{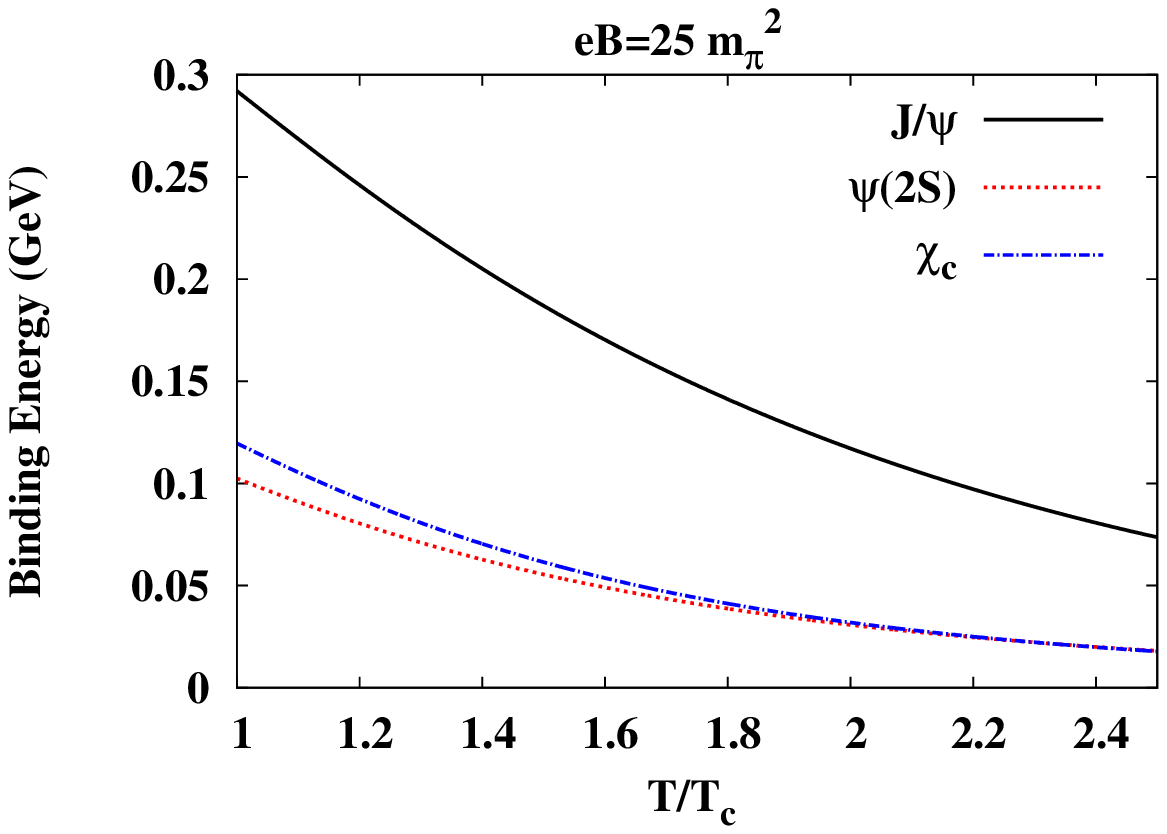}
\end{tabular}
\caption{The binding energies of quarkonia in pure thermal 
medium (left) and thermal medium in presence of magnetic field (right).}
\end{center}
\label{fig1}
\end{figure}   

\begin{figure}[h]
\begin{center}
\includegraphics[width=6.5cm,height=6.5cm]{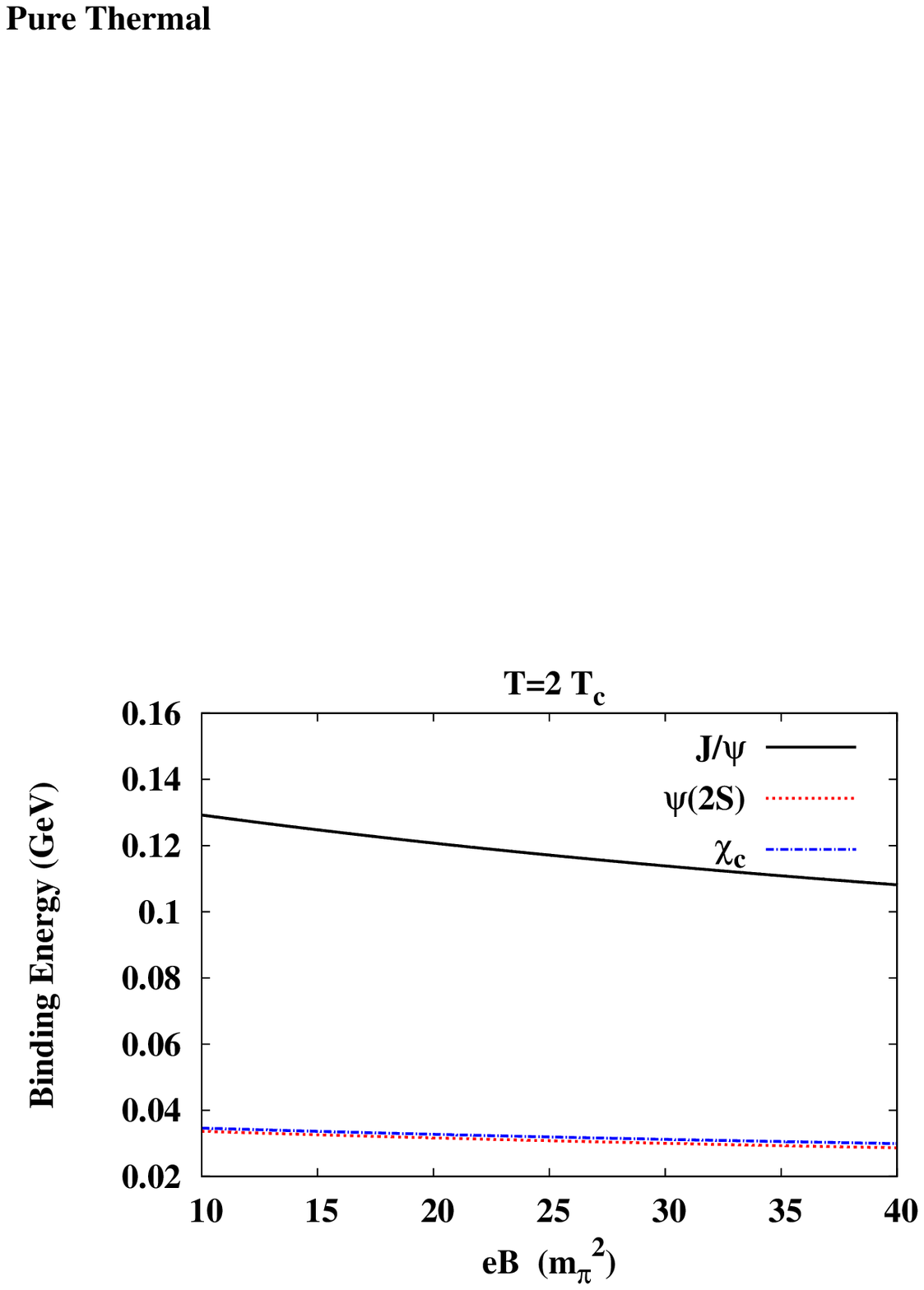}
\caption{Variation of binding energies with the magnetic field}
\end{center}
\label{fig1}
\end{figure}

\clearpage
\subsection{Thermal Width and Dissociation of Quarkonia}
Using the first-order 
perturbation theory, the width ($\Gamma$) has been evaluated numerically by 
folding the eigenstates of a specific quarkonium state in the deconfined 
medium in the presence of magnetic field 
\begin{eqnarray}
\Gamma =-2\int_0^\infty \rm{Im}~V(r;B,T) |\Phi_i(r)|^2 d\tau.
\label{gammaT}
\end{eqnarray}
We have thus computed the width as a function of temperature in 
absence (presence) of magnetic field in the left
(right) panel of Figure 10, respectively. We have found that in pure thermal
medium (left panel) the width increases with the temperature faster than 
in the presence of strong magnetic field (right panel). However, the magnetic
field always enhances the width of the resonances (Figure 11).
\begin{figure}[h]
\begin{center}
\begin{tabular}{c c}
\includegraphics[width=6.5cm,height=6.5cm]{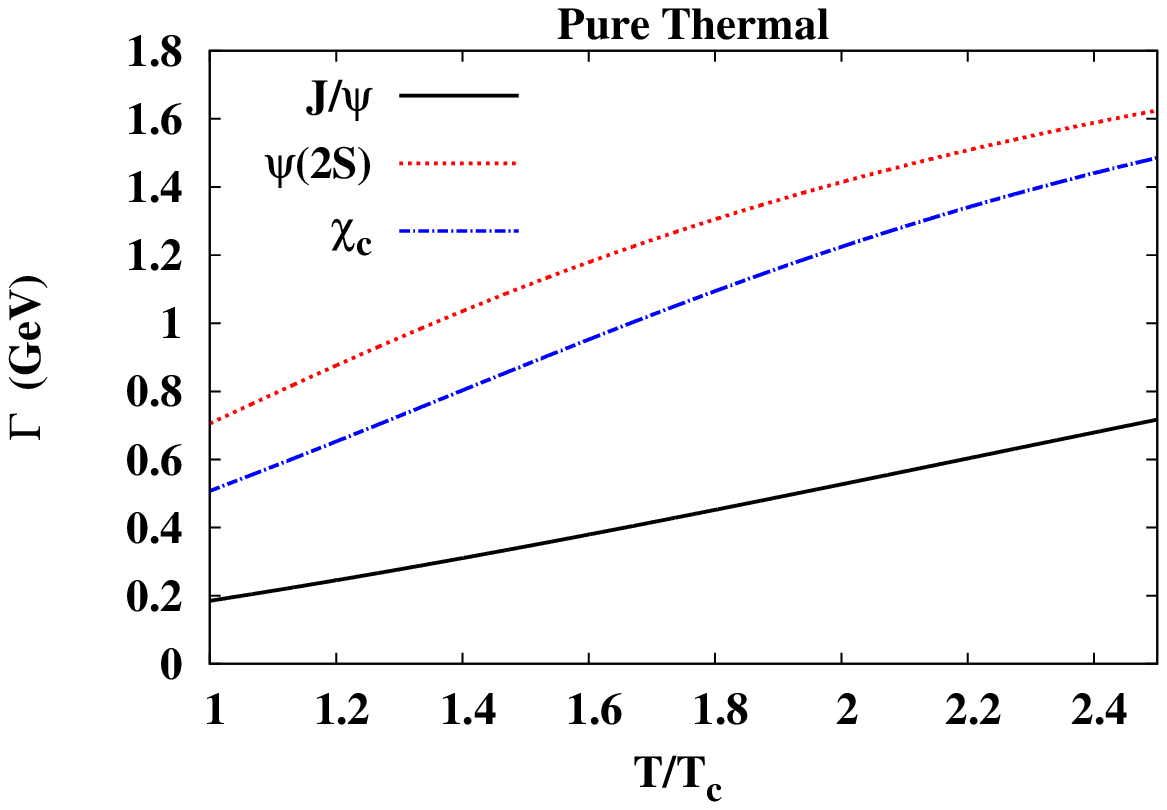}&
\includegraphics[width=6.5cm,height=6.5cm]{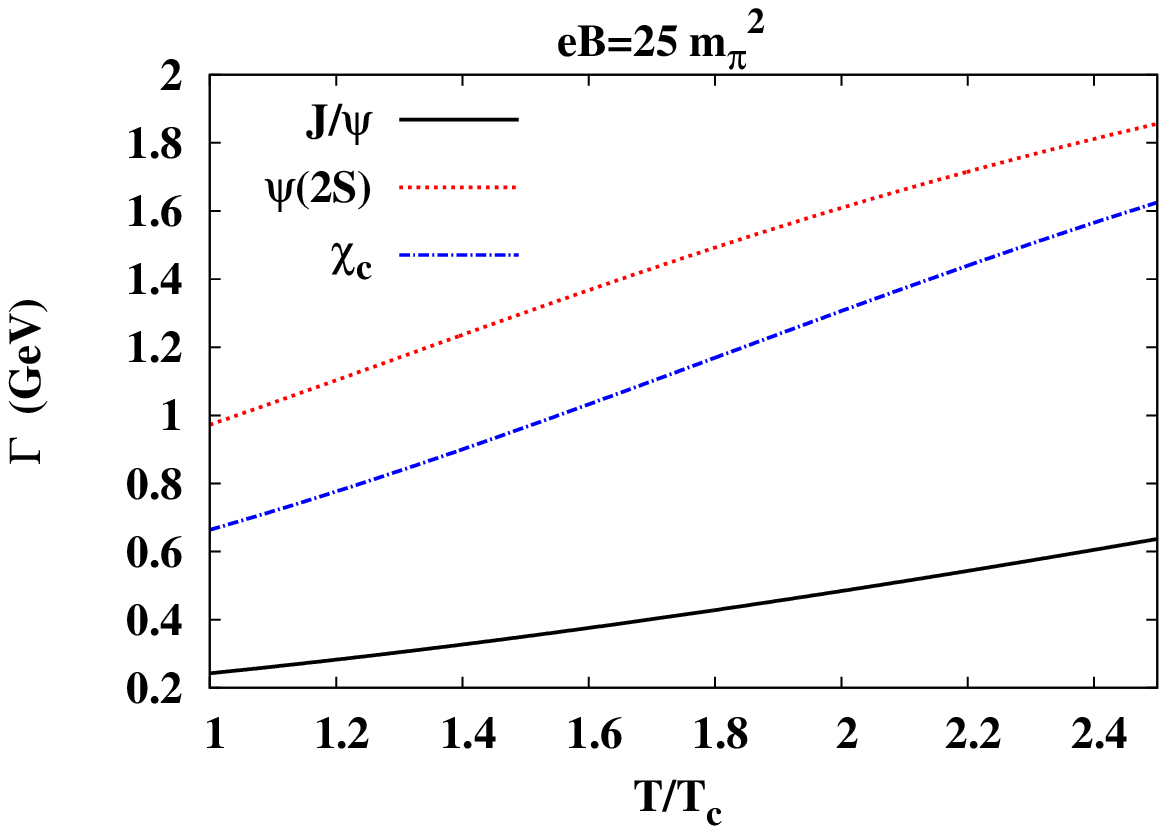}
\end{tabular}
\caption{Variation of the thermal widths with the temperature of 
the medium in absence (left) as well as presence (right) of
magnetic field}
\end{center}
\label{fig1}
\end{figure}

\begin{figure}[h]
\begin{center}
\begin{tabular}{c }
\includegraphics[width=6.5cm,height=6.5cm]{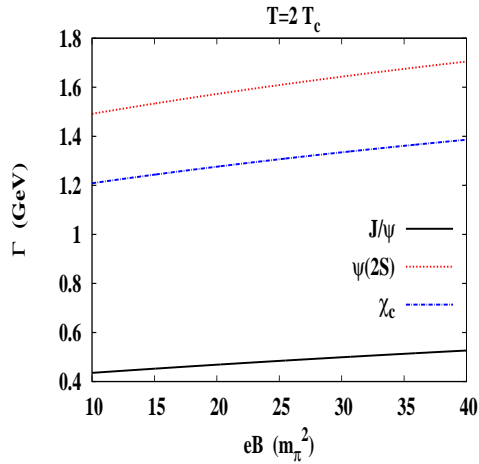}
\end{tabular}
\caption{Thermal widths of quarkonia is plotted as a function of
magnetic field}
\end{center}
\label{fig1}
\end{figure}

Having studied the change of properties of quarkonia in the presence
of magnetic field, we investigate now the effect of strong magnetic field 
on the dissociation of quarkonia from the conservative 
criterion on the width of the resonance in $\Gamma \ge 2 \rm{Re}~{\rm{B.E.}}$
\cite{Mocsy:PRL99'2007}. So we have first estimated the dissociation
temperatures of quarkonia in absence of magnetic field in the Table 1 and 
then did the same in presence of magnetic fields in 
Table 2. We found that the dissociation temperatures increase due to
the presence of strong magnetic field but with the further increase of 
magnetic field the dissociation temperatures decrease. {\em 
For example}, $J/\psi$'s and $\chi_c$'s are dissociated at higher temperatures
at 2 $T_c$ and 1.1 $T_c$ at a magnetic field $eB \approx 6 
~{\rm{and}}~ 4~m_\pi^2$, respectively, compared to the     
values 1.60 $T_c$ and 0.80$T_c$ in the absence of magnetic field,
respectively. However, the $J/\psi$
is dissociated at smaller temperatures, 1.8 $T_c$ and 1.5 $T_c$ for
higher magnetic fields, $eB=27~{\rm{and}}~68~m_\pi^2$, respectively.
Similarly for higher magnetic field, $eB=12~m_\pi^2$, $\chi_c$ gets
dissociated at the critical temperature.

\begin{table}[h]
 \begin{center}
 \begin{tabular}{|l|l|}
 \hline
 State & Dissociation Temperature\\ 
       & $T_D$~(in $T_c$) \\
 \hline
 \hline
 $J/\psi$ & 1.60\\
 \hline
 $\chi_c$ & 0.80\\
 \hline
 $\psi(2S)$ & 0.70\\
 \hline
 \end{tabular}
 \end{center}
 \caption{Dissociation temperature for thermal medium in absence of 
 magnetic field}
 \end{table}

\begin{table}[h]
\begin{center}
\begin{tabular}{|l|l|}
 \hline
 State & Dissociation Temperature (Magnetic field) \\
       & $T_D/T_c$ ($eB$~($m^2_\pi$))\\
 \hline
 \hline
 $J/\psi$ & 2.0 (6.50) \\
          & 1.8 (27.0) \\
          & 1.5 (68.0)\\
 \hline
 $\chi_c$ & 1.1 (3.7) \\
          & 1.0 (12) \\
 \hline          
 $\psi(2S)$ & $<1 (<m^2_\pi$)\\
 \hline
 \end{tabular}
 \end{center}
 \caption{Dissociation temperature for thermal medium in presence of 
 magnetic field} 
 \end{table}

\section{Conclusion}
The noncentral events in ultra-relativistic heavy-ion collisions
provide an opportunity to probe the properties of 
heavy quarkonia in the presence of a strong magnetic field. So we utilize 
this by calculating the bound state radii, binding energy, 
thermal width etc. of quarkonia by resummed perturbative thermal QCD 
in the presence of strong magnetic field, thereby studying 
the dissociation of quarkonia 
due to the Landau damping. For that purpose, using the Keldysh 
representation in real-time formalism, we have first calculated 
the real and imaginary parts of retarded gluon self-energy
for a deconfined medium in a strong magnetic field by 
thermalizing the Schwinger proper-time fermion propagator and then 
calculate the resummed retarded and symmetric propagators by
the Schwinger-Dyson equation. As a result, the Fourier components of 
both short and long distance components of 
$Q \bar Q$ interaction are being modified by the static limit
of resummed propagators and its inverse Fourier transform
gives rise the real and imaginary part of potential in coordinate
space. We have noticed that the long-distance
term is largely affected by the magnetic field than the shot-distance 
term, as a result the real part of potential becomes stronger and 
the imaginary part becomes larger than the medium in absence of magnetic field.

We have then studied the quarkonium dissociation by investigating its
properties by solving the Schrodinger equation numerically with potential 
derived to check
how the states and the probability distributions of quarkonia change
in the strong magnetic field. With the solutions of Schr\"{o}dinger equation
we have then calculated the average size, binding energy, thermal
width of resonances. We have found that the presence of strong magnetic field
causes the swelling for $J/\psi$ and squeezing for $\chi_c$. Similarly 
the binding decreases for $J/\psi$ and increases for $\chi_c$.
Moreover the 
binding energies decrease with the temperature of medium very slowly
due to the presence of magnetic field and for a given medium the binding
decreases with the increase in magnetic fields. On the other hand the
presence of magnetic field causes an increase the width of resonances
in a hot QCD medium.

The above observations on the change of the properties of quarkonia
in a strong magnetic field facilitate to study the dissociation of quarkonia 
due to the Landau damping and quantify 
the magnetic field at which a specific $Q \bar Q$ state 
excites to the continuum from the intersection of the magnetic field induced 
thermal width and the (twice) binding energy curve.
We have noticed that the presence of strong magnetic field increase the 
dissociation temperatures but it decreases with the further increase of 
magnetic field. {\em For example}, $J/\psi$'s and $\chi_c$'s 
are dissociated at higher temperatures
at 2 $T_c$ and 1.1 $T_c$ at a magnetic field $eB \approx 6 m_\pi^2$ and 
$4 m_\pi^2$, respectively, compared to the 
values 1.60 $T_c$ and 0.8 $T_c$ in the absence of magnetic field, respectively.

\section {Acknowledgement}
BKP is thankful to the CSIR (Grant No.03 (1407)/17/EMR-II),
Government of India for the financial assistance.


\end{document}